\documentclass[a4paper,fleqn,usenatbib]{mn2e}

\usepackage[english]{babel}
\usepackage{graphicx}
\usepackage{fleqn}
\usepackage{amsfonts}
\usepackage{amssymb}
\usepackage{amsmath}
\usepackage{booktabs}
\usepackage[flushleft]{threeparttable}

\newcommand{\aap}{A\&A}
\newcommand{\aj}{AJ}
\newcommand{\apj}{ApJ}
\newcommand{\apjs}{ApJs}
\newcommand{\apjl}{ApJL}

\newcommand{\icarus}{Icar}

\newcommand{\planss}{P\&SS}
\newcommand{\mnras}{MNRAS}
\newcommand{\nat}{Nature}

\renewcommand{\S}{Section}

\newcommand{\quadpaper}{HPAPZ15}

\title[Circumbinary planets in triples]{A triple origin for the lack of tight coplanar circumbinary planets around short-period binaries}
\author[Adrian S. Hamers, Hagai B. Perets and Simon F. Portegies Zwart]{Adrian S. Hamers$^{1}$\thanks{E-mail: hamers@strw.leidenuniv.nl}, Hagai B. Perets$^{2}$ and Simon F. Portegies Zwart$^{1}$\\
$^{1}$Leiden Observatory, Leiden University, PO Box 9513, NL-2300 RA Leiden, The Netherlands \\
$^{2}$Technion - Israel Institute of Technology, Haifa 32000, Israel}

\date{MNRAS 455, 3, 3180-3200 (2016) \\
Accepted 2015 October 20. Received 2015 October 20; in original form 2015 June 5}

\pubyear{2015}

\begin{document}
\label{firstpage}
\pagerange{\pageref{firstpage}--\pageref{lastpage}}
\maketitle

\begin{abstract}
Transiting circumbinary planets are more easily detected around short-period than long-period binaries, but none have yet been observed by {\it Kepler} orbiting binaries with periods shorter than seven days. In triple systems, secular Kozai-Lidov cycles and tidal friction (KLCTF) have been shown to reduce the inner orbital period from $\sim 10^4$ to a few days. Indeed, the majority of short-period binaries are observed to possess a third stellar companion. Using secular evolution analysis and population synthesis, we show that KLCTF makes it unlikely for circumbinary transiting planets to exist around short-period binaries. We find the following outcomes. (1) Sufficiently massive planets in tight and/or coplanar orbits around the inner binary can quench the KL evolution because they induce precession in the inner binary. The KLCTF process does not take place, preventing the formation of a short-period binary. (2) Secular evolution is not quenched and it drives the planetary orbit into a high eccentricity, giving rise to an unstable configuration, in which the planet is most likely ejected from the system. (3) Secular evolution is not quenched but the planet survives the KLCTF evolution. Its orbit is likely to be much wider than the currently observed inner binary orbit, and is likely to be eccentric and inclined with respect to the inner binary. These outcomes lead to two main conclusions: (1) it is unlikely to find a massive planet on a tight and coplanar orbit around a short-period binary, and (2) the properties of circumbinary planets in short-period binaries are constrained by secular evolution.
\end{abstract}


\section{Introduction}
\label{sect:introduction}
An increasing number of transiting circumbinary planets around solar-type main-sequence (MS) binary stars are being discovered by the {\it Kepler} mission. Currently, 10 such planets are known (see Table \ref{table:observed_systems}; top rows). The orbital periods of the stellar binaries in the systems discovered so far have a mean value of $20.4$ d; the shortest period is 7.45 d (Kepler 47; \citealt{orosz_ea_12b}). In contrast, the {\it Kepler} eclipsing binaries {\it without} circumbinary planets typically have shorter orbital periods, with a mean period of 2.8 d. There exists a bias for detecting {\it more} transiting circumbinary planets around shorter-period binaries because (1) the shorter period results in more transits in a given amount of time, and (2) the binary precession period is only a few years long, resulting in intersecting binary and planet orbits (as seen in the plane of the sky) during the {\it Kepler} mission \citep{martin_triaud_15}. Because of this bias, many circumbinary planets are expected to have been observed around short-period {\it Kepler} eclipsing binaries. However, so far, none have been found. 

If the apparent lack of (nearly) coplanar circumbinary planets around close binaries is intrinsic and not related to (yet unknown) observational bias(es), then this raises the question of its origin. The following two observations suggest that this origin is related to the (MS) evolution of solar-type triple star systems. 
\begin{itemize}
\item The fraction of tertiary companions to spectroscopic binaries is a strong function of the inner period, increasing from 0.34 for inner periods $P_\mathrm{A}>12\,\mathrm{d}$ to 0.96 for $P_\mathrm{A}<3 \, \mathrm{d}$ \citep{tokovinin_ea_06}. For the {\it Kepler} eclipsing binaries in particular, the tertiary fraction for inner binaries with periods $\lesssim 3 \, \mathrm{d}$ is $\sim 20 \%$ \citep{rappaport_ea_13,conroy_ea_14}. The latter studies are limited (by observing time) to triples with outer periods $\lesssim 3 \, \mathrm{yr}$. The complete tertiary fraction for these {\it Kepler} binaries, i.e. including any outer period, is likely much larger. 
\item In the survey of \citet{tokovinin_14a}, a peak is found in the (inner) period distribution around 3 d. As shown by \citet{fabrycky_tremaine_07,naoz_fabrycky_14}, this peak in the period distribution can be explained by the combination of the secular gravitational torque of the tertiary companion and tidal friction in the inner binary. The former, through Kozai-Lidov (KL) oscillations \citep{lidov_62,kozai_62}, can excite the inner binary orbit to high eccentricity. Consequently, tides are much more effective, leading to tidal dissipation and shrinkage of the orbit, and, eventually, to a tight and (nearly) circular orbit. In this scenario, the precursors of the binaries affected by KL cycles with tides have orbital periods ranging roughly between $10$ and $10^4$ d.
\end{itemize}

\begin{figure}
\center
\includegraphics[scale = 0.45, trim = 10mm 0mm 0mm 0mm]{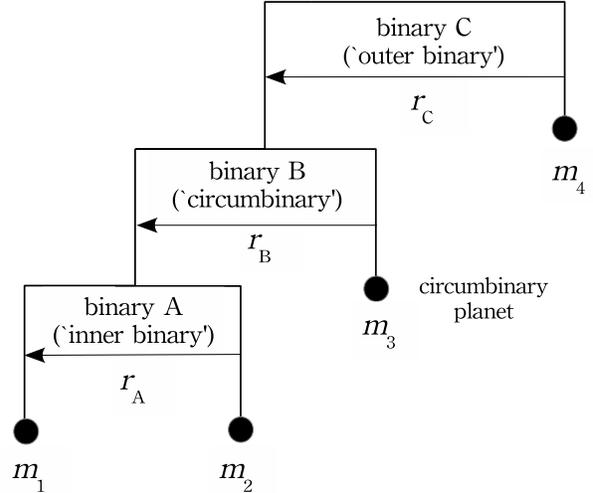}
\caption{\small A schematic depiction (not to scale) of the hierarchical configuration of the systems considered in this paper. The circumbinary planet ($m_3$) orbits the inner binary ($m_1$ and $m_2$); the outermost star ($m_3$), in the outer binary, orbits the centre of mass of the inner binary+planet system. }
\label{fig:definitions}
\end{figure}

The lower limit of 10 d of precursor binaries is right in the ballpark of the orbital periods of the currently observed {\it Kepler} systems with transiting circumbinary planets. Also, there are a number of observed wider systems orbited by at least one circumbinary planet; some of them are listed in Table \ref{table:observed_systems} (bottom rows). Therefore, a large fraction of these precursor binaries might have a circumbinary planet in a stable orbit in-between the `inner' and `outer' {\it stellar} orbits (henceforth, we refer to the inner and outer stellar binaries simply as the `inner' and `outer' binaries; see also Fig. \ref{fig:definitions}). 

Here, we show that in these stellar triples with a circumbinary planet around the inner binary, the planet can strongly affect the secular orbital evolution of the inner binary compared to the situation without a planet, provided that certain conditions are met (see below). This `shielding' effect arises from a quenching of the eccentricity oscillations in the inner binary induced by the torque of the outer binary, because of precession induced from the circumbinary planet. The shielding effect is similar to the quenching of KL oscillations in stellar triples because of additional sources of orbital precession (notably, precession associated with general relativity, tidal bulges and/or stellar rotation). In some cases, however, resonances can occur that {\it enhance} the eccentricity oscillations in the inner binary (such resonances can also occur in isolated triples in conjunction with general relativity; \citealt{naoz_ea_13b}). 

Typically, shielding of KL oscillations by the presence of the planet is only effective for massive planets that are in a (nearly) coplanar and tight orbit with respect to the inner binary. This implies the following two scenarios for the secular evolution of the system.
\begin{itemize}
\item If the precursor binary is orbited by a massive planet in a tight and coplanar orbit, then planet shielding can prevent the binary from shrinking through the combined effects of KL cycles and tidal friction (KLCTF).
\item On the other hand, if the planet is of low mass and in an inclined and wide orbit with respect to the inner binary, then shielding is typically ineffective, and the binary can shrink through KLCTF to become a short-period binary.
\end{itemize}
Assuming that short-period binaries are produced through KLCTF, then this implies an intrinsic lack of massive planets in tight and coplanar circumbinary orbits (i.e. strongly shielding planets), whereas far-away and inclined low-mass circumbinary planets (i.e. weakly shielding planets) could still be abundant. This is consistent with the current null-detections of transiting {\it Kepler} circumbinary planets around short-period binaries. 

The goal of this paper is to quantify the above argument. The organization is as follows. In \S\,\ref{sect:methods}, we briefly describe our methods and assumptions. In \S\,\ref{sect:shielding}, we study the secular gravitational dynamics of stellar triples with a circumbinary planet, and we quantify the conditions when the planet can affect the inner binary. In order to evaluate the effect of circumbinary planets in a population of triples in the field, we carry out a population synthesis study in \S\,\ref{sect:pop_syn}. We discuss our results in \S\,\ref{sect:discussion}, where we also describe, using direct $N$-body integrations, the fate of circumbinary planets that become unstable because of secular evolution. We conclude in \S\,\ref{sect:conclusions}.

We remark that nearing the completion of this paper, we became aware of two similar and independent studies on the lack of circumbinary planets around short-period binaries in stellar triples by \citet{martin_mazeh_fabrycky_15} and \citet{munoz_lai_15}.

\section{Methods and assumptions}
\label{sect:methods}
We model the system of a circumbinary planet orbiting the inner binary in a stellar triple system as a hierarchical quadruple system in the `triple-single' configuration, which we studied previously in \citet{hamers_ea_15} (hereafter \quadpaper). In the latter work, the Hamiltonian of the system was derived and expanded in terms of the ratios of the three binary separations $r_\mathrm{A}$, $r_\mathrm{B}$ and $r_\mathrm{C}$, where, by assumption, $r_\mathrm{C} \gg r_\mathrm{B} \gg r_\mathrm{A}$. In the current context, binary A corresponds to the inner binary, binary B to the orbit of the planet around the inner binary, and binary C to the orbit of the tertiary star around the centre of mass of the inner binary+planet system (evidently, the latter nearly coincides with the centre of mass of the inner binary). For consistency with \quadpaper, we use indices 1 and 2 to denote quantities associated with the inner binary primary and secondary, respectively, index 3 for the planet, and index 4 for the tertiary star. 

A schematic depiction of our configuration is shown in Fig. \ref{fig:definitions}.

In \quadpaper, the orbit-averaged Hamiltonian was derived, and a numerical algorithm was developed within the \textsc{AMUSE} framework \citep{portegies_zwart_ea_13,pelupessy_ea_13} to solve the resulting equations of motion. Post-Newtonian (PN) dynamics at the 1PN order are taken into account as described by equation (9) of \quadpaper. 

Here, we extended this algorithm by also including the effects of tidal friction within the inner binary. Gravitational perturbations by the planet and the tertiary star with regards to the inner binary tidal evolution were ignored. We adopted the equilibrium tide model \citep{eggleton_ea_98}, in which it is assumed that each star quasi-hydrostatically adjusts its shape to the time-varying perturbing potential of its companion. Following \citet{barker_ogilvie_09}, we adopted a constant tidal quality factor $Q$ related to the mean motion $n$ and the tidal lag time $\tau$ via $Q=1/(n \tau)$. Below, instead of $Q$, we use the directly related quantity $Q'\equiv 3Q/(2k)$ \citep{barker_ogilvie_09}, where $k$ is the second-order potential Love number. Typical values of $Q'$ for solar-type stars, as inferred from observations, are $Q'\sim 5.5\times 10^5-2\times 10^6$ \citep{meibom_mathieu_05,ogilvie_lin_07}. 

A constant $Q$ implies that $\tau = 1/(nQ) = P/(2\pi Q)$ scales with the orbital period, and, therefore, $\tau$ effectively decreases as the orbital period decreases due to tidal friction. This may not give an entirely accurate description; e.g. \citet{socrates_ea_12a} and \citet{socrates_ea_12b} show that with a number of simplifying assumptions, $\tau$ is constant, and the results of \citet{hansen_10} suggest that $Q'$ is not constant and dependent on the semimajor axis (cf. equation 13 of \citealt{hansen_10}). Nevertheless, as argued by \citet{barker_ogilvie_09}, given the current uncertainties in the underlying tidal dissipation mechanisms and, therefore, the efficiency of tidal dissipation, the assumption of a constant $Q$ is useful for studying the general effects of tidal friction. 

The equilibrium tide model is included in our algorithm by adding the relevant terms $\mathrm{d} \boldsymbol{e}_\mathrm{A}/\mathrm{d}t|_\mathrm{TF}$, $\mathrm{d} \boldsymbol{h}_\mathrm{A}/\mathrm{d}t|_\mathrm{TF}$, $\mathrm{d} \boldsymbol{\Omega}_1/\mathrm{d}t|_\mathrm{TF}$ and $\mathrm{d} \boldsymbol{\Omega}_2/\mathrm{d}t|_\mathrm{TF}$ in the equations of motion, as given by equations (A7)-(A15) of \citet{barker_ogilvie_09}. Here, $\boldsymbol{e}_k$ and $\boldsymbol{h}_k$ are the eccentricity vector and the specific angular momentum vector of orbit $k$, respectively, and $\boldsymbol{\Omega}_k$ is the spin angular momentum vector of star $k\in \{1,2\}$. Throughout this paper, the initial spin vectors $\boldsymbol{\Omega}_k$ are assumed to be parallel with the inner orbital angular momentum vector $\boldsymbol{h}_\mathrm{A}$.

Our algorithm, that models the orbital evolution, is coupled within \textsc{AMUSE} with the stellar evolution code \textsc{SeBa} \citep{portegies_zwart_verbunt_96,toonen_ea_12}, which is also interfaced within \textsc{AMUSE}. We use the latter code to compute the masses and radii during in the integration. In this work, we focus on low-mass MS stars. Therefore, the time-dependence of the latter quantities is weak for time-scales less than a Hubble time. We remark, however, that this is no longer the case for post-MS evolution.

\begin{table*}
\begin{threeparttable}
\begin{tabular}{lccccc}
\S & \ref{sect:shielding:kepler} & \ref{sect:shielding:tides} & \ref{sect:discussion:unstable} & \ref{sect:pop_syn} & \\
\toprule
$m_1/\mathrm{M}_\odot$ & 1 & 1 & 0.83 & 0.5-1.2 \tnote{a} \\
$m_2/\mathrm{M}_\odot$ & 0.5 & 0.5 & 0.50 & 0.1-1.2 \tnote{b} \\
$m_3/M_\mathrm{J}$ & 0.01-1 &0.001-3.16 & 0.01-1 & 0.01-1 \\
$m_4/\mathrm{M}_\odot$ & 1 & 1 & 0.73 & 0.1-1.2 \tnote{c} \\
$a_\mathrm{A}/\mathrm{AU}$ & 0.2 & 1 & 0.20 & 0.03-2.0 \tnote{d} \\ 
$a_\mathrm{B}/\mathrm{AU}$ & 1-5 & 10-50 & 1.09 - 1.22 & 0.14- \\
& & & & 3.1$\times 10^2$ \tnote{e}\\ 
$a_\mathrm{C}/\mathrm{AU}$ & 20 & 160 & 7.73 & 1.5- \\
& & & & 1.8$\times 10^4$ \tnote{d} \\
$e_\mathrm{A}$ & 0.01 & 0.01 & 0.40 & 0.01-0.95 \tnote{f} \\
$e_\mathrm{B}$ & 0.01 & 0.01 & 0.01 & 0.01 \\
$e_\mathrm{C}$ & 0.01 & 0.01 & 0.30 & 0.01-0.95 \tnote{f} \\
$i_\mathrm{A}/{}^\circ$ & 0.01 & 0.01 & 0.01 & 0.01 \\
$i_\mathrm{B}/{}^\circ$ & 0.01-180 & 0.01-180 & 0.01-180 & 0.01-180 \\
$i_\mathrm{C}/{}^\circ$ & 75 & 85 & 73.7 & 0.01-180 \tnote{g} \\
$\omega_\mathrm{A}/{}^\circ$ & 0.01 & 0.01 & 287.7 & 0.01-180 \tnote{h} \\
$\omega_\mathrm{B}/{}^\circ$ & 0.01 & 0.01 & 281.0 & 0.01-180 \tnote{h} \\
$\omega_\mathrm{C}/{}^\circ$ & 0.01 & 0.01 & 51.6 & 0.01-180 \tnote{h} \\
$\Omega_\mathrm{A}/{}^\circ$ & 0.01 & 0.01 & 166.1 & 0.01-180 \tnote{h} \\
$\Omega_\mathrm{B}/{}^\circ$ & 0.01 & 0.01 & 42.6 & 0.01-180 \tnote{h} \\
$\Omega_\mathrm{C}/{}^\circ$ & 0.01 & 0.01 & 340.1 & 0.01-180 \tnote{h} \\
$P_{\mathrm{spin},k}/\mathrm{d}$ & N/A & 10 & 10 & 10 \\
$Q'_k/10^6$ & N/A & 1 & 1.9 & 0.55-2 \tnote{h} \\
$r_{\mathrm{g},k}$ & N/A & 0.08 & 0.08 & 0.08 \\
\bottomrule
\end{tabular}
\begin{tablenotes}
            \item[a] Salpeter distribution $\mathrm{d}N/\mathrm{d}m_1 \propto m_1^{-2.35}$.
            \item[b] Sampled from $m_2 = q_\mathrm{in} m_1$, where $0<q_\mathrm{in}<1$ is linearly distributed. 
            \item[c] Sampled from $m_4 = q_\mathrm{out} (m_1+m_2)$, where $0<q_\mathrm{out}<1$ is linearly distributed. 
            \item[d] Lognormal distribution in the orbital period $P_k$ with mean $\log_{10}(P_k/\mathrm{d}) = 5.03$, standard deviation $\sigma_{\log_{10}(P_k/\mathrm{d})} = 2.28$ and range $-2 < \log_{10} (P_k/\mathrm{d}) < 10$. Triples are subject to stability constraints, and the inner binary should not merge in the absence of a planet.
            \item[e] Linearly sampled for each triple, with lower limit $a_\mathrm{B,l} = 1.5 \, a_\mathrm{B,crit,AB}$, where $a_\mathrm{B,crit,AB}$ is the critical semimajor axis for dynamical stability of the planet in a coplanar orbit around the inner binary, and which is adopted from the fitting formula given by \citep{holman_wiegert_99}. The upper limit is $a_\mathrm{B,u} = 0.9\,a_\mathrm{B,crit,BC}$, where $a_\mathrm{B,crit,BC}$ is the largest possible value of $a_\mathrm{B}$ for dynamical stability with respect to the orbit of the tertiary star, estimated by applying the \citealt{mardling_aarseth_01} criterion to the BC pair, with an `outer' mass ratio of $q_\mathrm{out} = m_4/(m_1+m_2+m_3)$.
            \item[f] Sampled from a Rayleigh distribution, $\mathrm{d} N/\mathrm{d} e_k \propto e_k \exp(-\beta e_k^2)$, with rms $\langle e_k^2 \rangle^{1/2} = \beta^{-1/2} = 0.33$ \citep{raghavan_ea_10}.
            \item[g] Sampled from a linear distribution in $\cos(i_\mathrm{C})$, with $-1<\cos(i_\mathrm{C})<1$.
            \item[h] Sampled from a linear distribution.
\end{tablenotes}
\caption{ Initial conditions for the systems in \S\,\ref{sect:shielding:kepler} (first column, cf. Figs \ref{fig:maximum_eccentricity_test02}-\ref{fig:maximum_eccentricity_large_grid_test02ai3}), for the systems in \S\,\ref{sect:shielding:tides} (second column, cf. Figs \ref{fig:tidal_friction_elements_time_paper_test04Lm.eps} and \ref{fig:tidal_friction_test04}), for the systems in \S\,\ref{sect:discussion:unstable} (third column, cf. Fig. \ref{fig:nbody_check_dynamical_stability}), and for the population synthesis in \S\,\ref{sect:pop_syn} (fourth column). Note that for the third column, the systems are integrated with the secular code until the orbit of the planet intersects with the inner binary; subsequent evolution is carried out using direct $N$-body integration (cf. \S\,\ref{sect:discussion:unstable}). The orbital angles (inclinations $i_k$, arguments of pericentre $\omega_k$ and longitudes of the ascending node $\Omega_k$) are defined with respect to an arbitrary, but fixed reference frame; see equation (6) of \quadpaper\, for the relation between these angles and the orbital vectors. The spin periods $P_{\mathrm{spin},k}$, tidal quality factors $Q'_k$ and radii of gyration $r_{\mathrm{g},k}$ apply to the inner binary stars and are used for tidal evolution. }
\label{table:init_cond1}
\end{threeparttable}
\end{table*}

\begin{figure*}
\center
\includegraphics[scale = 0.68, trim = 20mm 0mm 0mm 0mm]{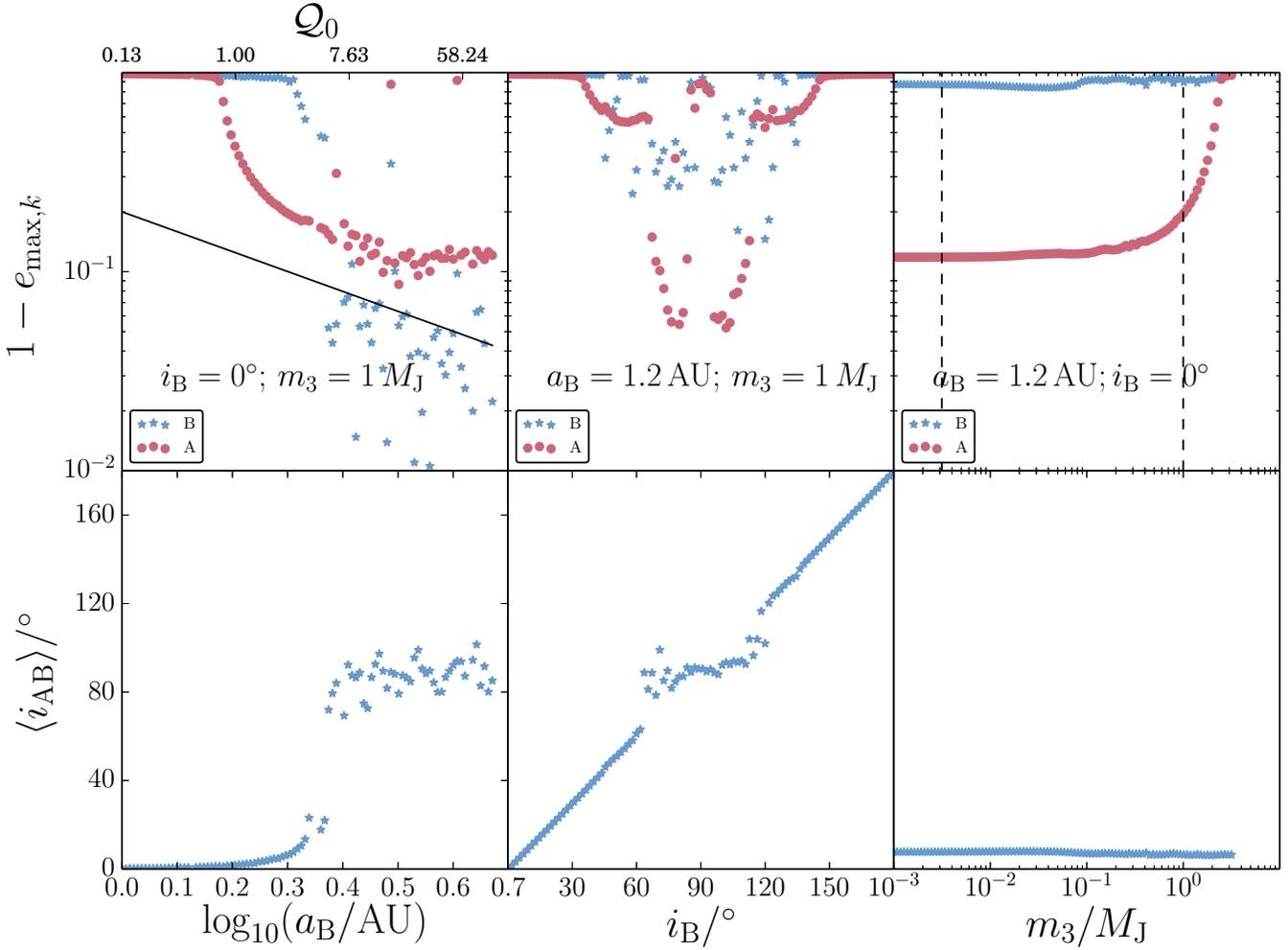}
\caption{\small Top row: maximum eccentricities in the inner binary (red bullets) and circumbinary orbit (blue stars) for the integrations of \S\,\ref{sect:shielding:kepler} (cf. the left column of Table \ref{table:init_cond1}), as a function of the initial $a_\mathrm{B}$ (left column), $i_\mathrm{B}$ (middle column) and $m_3$ (right column). In the left column $i_\mathrm{B}=0^\circ$ and $m_3=1\,M_\mathrm{J}$; in the middle column $a_\mathrm{B}=1.2\,\mathrm{AU}$ and $m_3=1\,M_\mathrm{J}$; and in the right column $a_\mathrm{B}=1.2\,\mathrm{AU}$ and $i_\mathrm{B}=0^\circ$. The corresponding value of $\mathcal{Q}_0$ (cf. equation~\ref{eq:Q0}) is indicated in the top left panel. In the bottom row, the average inclination between the inner and circumbinary orbits is shown. The two black vertical dashed lines in the top right panel indicate Earth and Jupiter masses, respectively. }
\label{fig:maximum_eccentricity_test02}
\end{figure*}

\begin{figure*}
\center
\includegraphics[scale = 0.62, trim = 32mm 0mm 0mm 0mm]{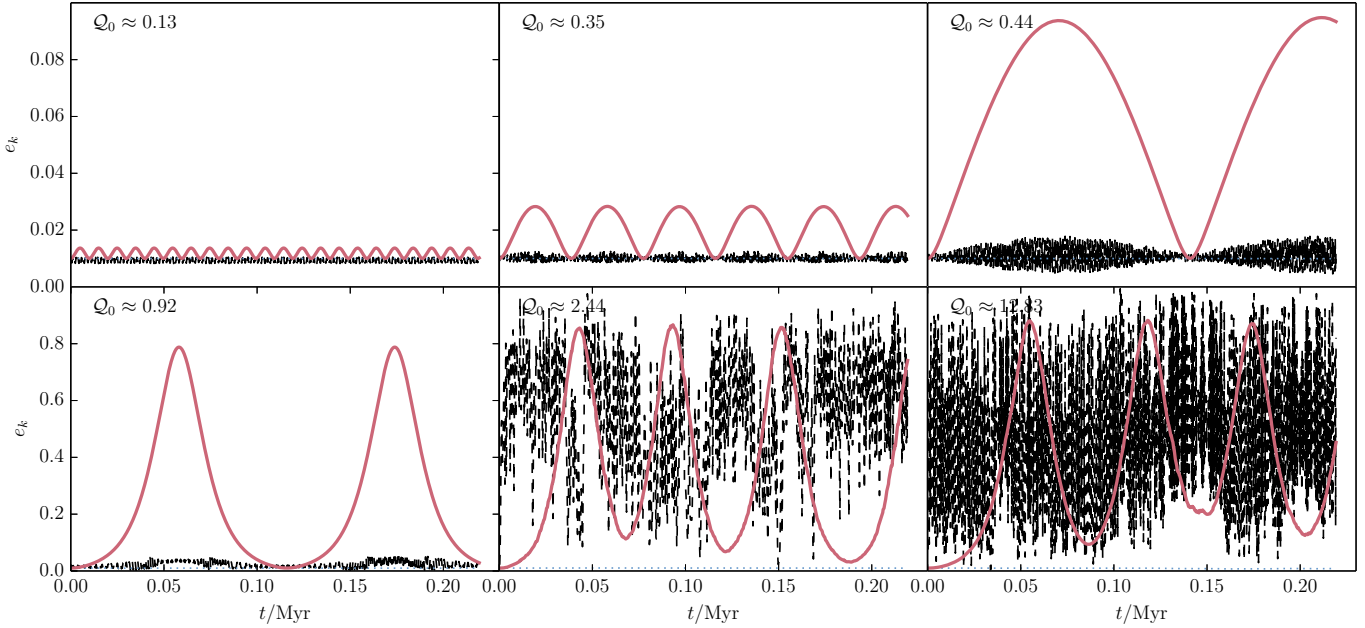}
\caption{\small Illustration of the time evolution of the eccentricities of the inner orbit and the circumbinary, for six systems corresponding to the top-left panel of Fig. \ref{fig:maximum_eccentricity_test02}. In each panel, the value of $\mathcal{Q}_0$ is indicated. }
\label{fig:elements_time_grid04NC}
\end{figure*}

\begin{figure}
\center
\includegraphics[scale = 0.45, trim = 0mm 0mm 0mm 0mm]{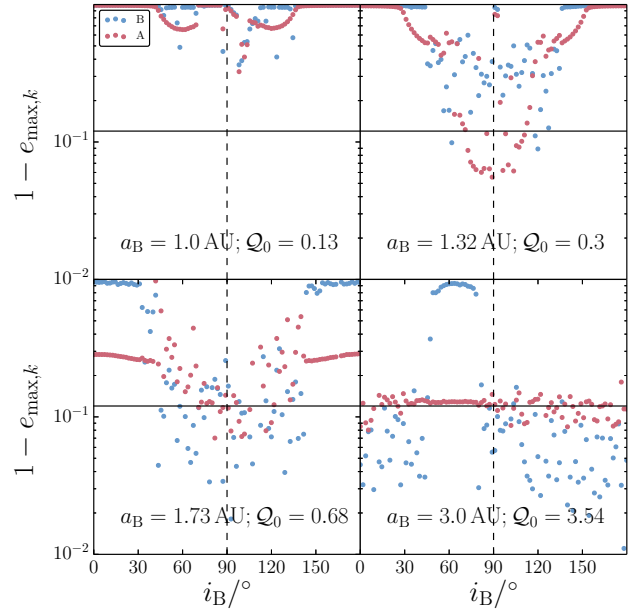}
\caption{\small Illustration of the joint dependence of the maximum eccentricities in binaries A and B on $a_\mathrm{A}$ and $i_\mathrm{B}$, as an extension of the top-middle panel of Fig. \ref{fig:maximum_eccentricity_test02}. In each panel, $1-e_{\mathrm{max},k}$ is plotted as a function of $i_\mathrm{B}$. The value of $a_\mathrm{B}$ is indicated for each panel; for all panels, $m_3=1\,M_\mathrm{J}$. The black solid horizontal lines show the value of $1-e_\mathrm{max,A}$ in the absence of the planet. }
\label{fig:maximum_eccentricity_large_grid_test02ai3}
\end{figure}

\section{The planet-shielding effect}
\label{sect:shielding}
Before studying the effect of circumbinary planets in a population of triples (cf. \S\,\ref{sect:pop_syn}), we first focus on a simplified case where the stellar triple system is kept fixed, and planets are `inserted' at stable orbits at various loci in-between the inner and outer binaries. The aim is to gain quantitative insight into the effect of the planetary orbit on the eccentricity oscillations and the tidal evolution in the inner binary. For simplicity, we here focus on triples in which the octupole-order terms are unimportant. Note, however, that the (non-cross) octupole-order terms are always included in the integrations (cf. \S\,\ref{sect:discussion:approximations}), and no restriction is made with respect to the importance on the octupole-order terms in \S\,\ref{sect:pop_syn}.

\subsection{{\it Kepler} transiting circumbinary systems}
\label{sect:shielding:kepler}
Here, we focus on systems similar to the {\it Kepler} transiting circumbinary systems. The inner semimajor axes in the latter are $a_\mathrm{A} \sim 0.1-0.2 \, \mathrm{AU}$ (cf. the top rows of Table \ref{table:observed_systems}), and relativistic precession is therefore important in these systems. As is well known from previous studies (e.g., \citealt{holman_touma_tremaine_97,blaes_ea_02,naoz_ea_13b}), general relativistic precession limits the range of semimajor axes and eccentricities of tertiary orbits for which the inner orbit eccentricity is excited. In general, an order-of-magnitude estimate of the KL time-scale for the binary pair $kl$ is given by
\begin{align}
P_{\mathrm{KL},kl} = \frac{P_l^2}{P_k} \frac{m_{k,\mathrm{p}} + m_{k,\mathrm{s}} + m_{l,\mathrm{s}}}{m_{l,\mathrm{s}}} \left ( 1-e_l^2 \right )^{3/2},
\label{eq:PK}
\end{align}
where $m_{k,\mathrm{p}}=m_1$, $m_{k,\mathrm{s}} = m_2$ and $m_{l,\mathrm{s}}=m_3$ in the case of $P_\mathrm{KL,AB}$, $m_{k,\mathrm{p}}=m_1+m_2$, $m_{k,\mathrm{s}} =m_3$ and $m_{l,\mathrm{s}}=m_4$ in the case of $P_\mathrm{KL,BC}$, and $m_{k,\mathrm{p}}=m_1$, $m_{k,\mathrm{s}} = m_2$ and $m_{l,\mathrm{s}}=m_4$ in the case of $P_\mathrm{KL,AC}$ (\citealt{innanen_ea_97}, \quadpaper, see also \citealt{antognini_15}). The (pairwise) 1PN time-scale for orbit $k$ is given by
\begin{align}
\label{eq:t_1PN}
t_{\mathrm{1PN},k} = \frac{1}{3} P_k \left(1-e_k^2 \right) \frac{a_k}{r_{\mathrm{g},k}},
\end{align}
where $r_{\mathrm{g},k} \equiv Gm_{\mathrm{tot},k}/c^2$, with $m_\mathrm{tot,A}=m_1+m_2$, $m_\mathrm{tot,B}=m_1+m_2+m_3$ and $m_\mathrm{tot,C}=m_1+m_2+m_3+m_4$, is the gravitational radius. Equating equation~(\ref{eq:PK}) as applied to the AC pair, to equation (\ref{eq:t_1PN}) as applied to binary A, we find that relativistic precession dominates in binary A if
\begin{align}
\nonumber a_\mathrm{C} &> \left(1 - e_\mathrm{C}^2 \right )^{-1/2} \left [ \frac{a_\mathrm{A}^4 c^2 \left(1-e_\mathrm{A}^2 \right) m_4(m_1+m_2+m_3+m_4)}{3G (m_1+m_2)^2 (m_1+m_2+m_4)} \right ]^{1/3} \\
&\approx 34 \, \mathrm{AU},
\end{align}
where for the numerical estimate we assumed $e_\mathrm{A}=e_\mathrm{C}=0.01$, $m_1 = 1 \, \mathrm{M}_\odot$, $m_2 = 0.5 \, \mathrm{M}_\odot$, $m_3 = 1 \, M_\mathrm{J}$ and $m_4 = 1 \, \mathrm{M}_\odot$. In the remainder of \S\,\ref{sect:shielding:kepler}, we shall assume these values for the numerical estimates.

An estimate of the lower limit on $a_\mathrm{C}$ for dynamical stability with respect to the inner binary (not yet taking into account the circumbinary planet, and ignoring the dependence on mutual inclination) is given by \citep{mardling_aarseth_01}
\begin{align}
a_\mathrm{C} > \frac{2.8 a_\mathrm{A}}{1-e_\mathrm{C}} \left[ \left ( 1+\frac{m_4}{m_1+m_2} \right ) \frac{1+e_\mathrm{C}}{\sqrt{1-e_\mathrm{C}}} \right ]^{2/5} \approx 0.67 \, \mathrm{AU}.
\end{align}
Here, we assume a fixed $a_\mathrm{C} = 20 \, \mathrm{AU}$. With this choice, the octupole parameter $\epsilon_\mathrm{oct,AC}$, defined as
\begin{align}
\label{eq:eps_oct}
\epsilon_\mathrm{oct,AC} \equiv \frac{m_1-m_2}{m_1+m_2} \frac{a_\mathrm{A}}{a_\mathrm{C}} \frac{e_\mathrm{C}}{1-e_\mathrm{C}^2},
\end{align}
is $\epsilon_\mathrm{oct,AC} \approx 1.7\times 10^{-5}$, indicating that octupole-order terms in the Hamiltonian expansion for the AC pair are unimportant \citep{lithwick_naoz_11,katz_ea_11,naoz_ea_13a,teyssandier_13,li_ea_14}. Assuming an initial mutual inclination of $i_\mathrm{AC}=75^\circ$ between binaries A and C, and with the inclusion of the 1PN terms, the maximum eccentricity attained in binary A, {\it without} a circumbinary planet, is $e_\mathrm{A,max} \approx 0.88$. 

As a next step, we include a circumbinary planet with mass $m_3$ in orbit around binary A with semimajor axis $a_\mathrm{B}<a_\mathrm{C}$, and with various mutual inclinations $i_\mathrm{AB}$. For simplicity, the circumbinary orbit is assumed to be initially nearly circular, i.e. $e_\mathrm{B}=0.01$. Assuming $i_\mathrm{AB}=0^\circ$, for dynamical stability with respect to the inner binary, the semimajor axis $a_\mathrm{B}$ must satisfy \citep{holman_wiegert_99}
\begin{align}
a_\mathrm{B} > a_\mathrm{A} f(e_\mathrm{A},\mu_\mathrm{A}) \approx 0.48 \, \mathrm{AU},
\end{align}
where $f(e_\mathrm{A},\mu_\mathrm{A})$ is a function of the inner binary eccentricity $e_\mathrm{A}$ and mass ratio $\mu_\mathrm{A} \equiv m_2/(m_1+m_2)$, given by equation (3) of \citet{holman_wiegert_99}. An estimate of the largest possible value of $a_\mathrm{B}$ for dynamical stability with respect to the outer orbit, binary C, ignoring dependencies on inclinations, is given by \citep{mardling_aarseth_01}
\begin{align}
\nonumber a_\mathrm{B} &< a_\mathrm{C} \frac{1-e_\mathrm{C}}{2.8} \left[ \left ( 1+\frac{m_4}{m_1+m_2+m_3} \right ) \frac{1+e_\mathrm{C}}{\sqrt{1-e_\mathrm{C}}} \right ]^{-2/5} \\
&\approx 5.8 \, \mathrm{AU}.
\end{align}
Note that, strictly speaking, the criterion of \citep{mardling_aarseth_01} only applies to three-body systems. However, by carrying out a number of $N$-body integrations with \textsc{Mikkola} \citep{mikkola_merritt_08} and \textsc{Sakura} \citep{ferrari_ea_14} within \textsc{AMUSE}, we find that circular circumbinary orbits with $a_\mathrm{B} = 5.8 \, \mathrm{AU}$ in the system of our choice are indeed dynamically stable on time-scales of a few multiple periods of the outer binary. 

Based on these estimates, we carried out integrations of the long-term secular dynamical evolution with our orbit-averaged algorithm (cf. \S\,\ref{sect:methods}), with $a_\mathrm{B}$ ranging between 1 and 5 AU, $i_\mathrm{AB}$ between $0$ and $180^\circ$, and $m_3$ between $10^{-2}$ and $10^{0} \, M_\mathrm{J}$. To simplify the interpretation, tidal evolution is not included in \S\,\ref{sect:shielding:kepler}. The integration time was set to $2P_\mathrm{KL,AC}$, i.e. the time-scale for two KL oscillations in the AC pair, without taking into account the effect of the planet. A comprehensive list of initial parameters is given in the left column of Table \ref{table:init_cond1}. 

In the left, middle and right panels of the first row of Fig. \ref{fig:maximum_eccentricity_test02}, we show the resulting maximum eccentricities in binaries A (red bullets) and B (blue stars) as a function of $a_\mathrm{B}$ (with $i_\mathrm{B}=0^\circ$ and $m_3=1\,M_\mathrm{J}$), $i_\mathrm{B}$ (with $a_\mathrm{B}=1.2\,\mathrm{AU}$ and $m_3=1\,M_\mathrm{J}$), and $m_3$ (with $a_\mathrm{B}=1.2\,\mathrm{AU}$ and $i_\mathrm{B}=0^\circ$), respectively. In the bottom row of the same figure, we show the average inclinations between orbits AB as a function of $a_\mathrm{B}$, $i_\mathrm{B}$ and $m_3$. In the top left panel, we also show an equivalent dependence on the parameter $\mathcal{Q}_0$, defined as the initial ratio of the KL time-scales for the AB and AC pairs,
\begin{align}
\label{eq:Q0}
\nonumber \mathcal{Q}_0 &\equiv \frac{P_\mathrm{KL,AB,0}}{P_\mathrm{KL,AC,0}} \\
\nonumber &= \left ( \frac{a_\mathrm{B}}{a_\mathrm{C}} \right )^3 \frac{m_4}{m_3} \frac{m_1+m_2+m_3+m_4}{m_1+m_2+m_4} \left ( \frac{1-e_\mathrm{B,0}^2}{1-e_\mathrm{C,0}^2} \right )^{3/2} \\
\nonumber &\approx \left ( \frac{a_\mathrm{B}}{a_\mathrm{C}} \right )^3 \frac{m_4}{m_3} \left ( \frac{1-e_\mathrm{B,0}^2}{1-e_\mathrm{C,0}^2} \right )^{3/2} \\
&\approx 0.13 \left ( \frac{a_\mathrm{B}}{1 \, \mathrm{AU}} \right )^3.
\end{align}
Here, in the third line, we neglected the planet mass compared to the sum of all stellar masses. We note that $\mathcal{Q}_0$ is closely related to the quantity $\mathcal{R}_0$ defined by \quadpaper, and is given by
\begin{align}
\label{eq:R0}
\nonumber \mathcal{R}_0 &\equiv \frac{P_\mathrm{KL,AB,0}}{P_\mathrm{KL,BC,0}} \\
\nonumber &= \left ( \frac{a_\mathrm{B}^3}{a_\mathrm{A} a_\mathrm{C}^2} \right )^{3/2} \left( \frac{m_1+m_2}{m_1+m_2+m_3} \right )^{1/2} \frac{m_4}{m_3} \left ( \frac{1-e_\mathrm{B,0}^2}{1-e_\mathrm{C,0}^2} \right )^{3/2} \\
\nonumber &\approx  \left ( \frac{a_\mathrm{B}^3}{a_\mathrm{A} a_\mathrm{C}^2} \right )^{3/2} \frac{m_4}{m_3} \left ( \frac{1-e_\mathrm{B,0}^2}{1-e_\mathrm{C,0}^2} \right )^{3/2} \\
&\approx \mathcal{Q}_0 \left ( \frac{a_\mathrm{B}}{a_\mathrm{A}} \right )^{3/2}.
\end{align}

If $i_\mathrm{B}=0^\circ$ and $\mathcal{Q}_0 \gg 1$, $e_\mathrm{max,A}\approx 0.88$ is the same as in the case without the planet (cf. the top-left panel of Fig. \ref{fig:maximum_eccentricity_test02}). In this limit, the torque of the outer orbit on the inner orbit (for which $P_\mathrm{KL,AC}^{-1}$ is a proxy) is much larger than the torque of the circumbinary orbit on the inner orbit (for which $P_\mathrm{KL,AB}^{-1}\ll P_\mathrm{KL,AC}^{-1}$ is a proxy). The circumbinary orbit therefore only has a small perturbative effect on the inner orbit, and the maximum eccentricity is not affected. If $\mathcal{Q}_0 \gg 1$, then also $\mathcal{R}_0 \gg 1$ (cf. equation~\ref{eq:R0}). This implies that the torque of binary B is not strong enough to maintain coplanarity between binaries A and B, as is shown in the bottom-left panel of Fig. \ref{fig:maximum_eccentricity_test02}. Effectively, binary B decouples from binary A, and the torque of the outer orbit results in high-amplitude eccentricity oscillations in binary B. 

The latter oscillations have large enough eccentricities for the circumbinary orbit to intersect with the inner binary, i.e. $r_\mathrm{p,B} = a_\mathrm{B}(1-e_\mathrm{B}) < r_\mathrm{a,A} = a_\mathrm{A}(1+e_\mathrm{A})$. This is the case for all points below the black solid line in the top-left panel of Fig. \ref{fig:maximum_eccentricity_test02}. For the purposes of this section, no stopping conditions were imposed on the eccentricities during the integrations. However, in reality, the circumbinary orbit is dynamically unstable for points below the black solid line. This is borne out by integrating some of the systems below this line with direct $N$-body integration, where the orbital parameters are set to correspond to the moment of maximum eccentricity of binary B. We investigate the possible outcomes for these cases of dynamical instability further in \S\,\ref{sect:discussion:unstable}.

On the other hand, if $i_\mathrm{B}=0^\circ$ and $\mathcal{Q}_0 \ll 1$, $e_\mathrm{max,A}\approx e_\mathrm{A,0}$. In this case, the net torque on the inner orbit is dominated by the circumbinary orbit. Consequently, the inner orbit precesses much more rapidly compared to the case without the planet, and this results in a quenching of the KL eccentricity oscillations otherwise induced by the outer orbit. Furthermore, the period of the latter oscillations is substantially reduced, as is illustrated in Fig. \ref{fig:elements_time_grid04NC}, where we show the time evolution of $e_\mathrm{A}$ and $e_\mathrm{B}$ for several values of $\mathcal{Q}_0$, assuming $i_\mathrm{B}=0^\circ$ and $m_3=1\,M_\mathrm{J}$. Note that the inclination between orbits A and B remains zero in this case (cf. the bottom-left panel of Fig. \ref{fig:maximum_eccentricity_test02}), and that octupole-order terms are not dominant, even for our smallest value $a_\mathrm{B} = 1 \, \mathrm{AU}$ (for $a_\mathrm{B} = 1 \, \mathrm{AU}$, $\epsilon_\mathrm{oct,AB} \approx 6.7\times 10^{-4}$). Therefore, the circumbinary orbit, on its own, does not induce eccentricity oscillations in binary A. 

Whenever the circumbinary orbit becomes unstable, the planet could collide with one of the stars, or be ejected from the system (see also \S\,\ref{sect:discussion:unstable}). Evidently, in either case, the planet can no longer shield the inner binary from KL-eccentricity oscillations induced by the outer orbit. Typically, this occurs whenever the circumbinary orbit is relatively close to the outer orbit ($\mathcal{Q}_0\gtrsim 1$), and is therefore strongly affected by the latter orbit's torque. In the same regime, however, shielding is ineffective. This implies that dynamical instability does not necessarily rule out the possibility for shielding by the planet. In the top-left panel of Fig. \ref{fig:maximum_eccentricity_test02}, this is indeed the case: the circumbinary orbits become unstable for $\mathcal{Q}_0 \gtrsim 1$, whereas shielding only occurs for $\mathcal{Q}_0 \lesssim 1$.

The dependence of the maximum eccentricities on $i_\mathrm{B}$ is shown in the top middle panel of Fig. \ref{fig:maximum_eccentricity_test02} for $a_\mathrm{B}=1.2 \, \mathrm{AU}$. There is a strong dependence of the planet's shielding ability on $i_\mathrm{B}$. For coplanar (either prograde or retrograde) orbits, $e_\mathrm{A,max}$ is close to the initial value $e_\mathrm{A,0}$. As the initial orbit of the planet becomes more inclined with respect to the inner binary, shielding becomes less efficient. Interestingly, for $i_\mathrm{B}$ close to $90^\circ$, shielding is again {\it more} efficient.

The nature of the dependence on $i_\mathrm{B}$ also strongly depends on $a_\mathrm{B}$. In Fig. \ref{fig:maximum_eccentricity_large_grid_test02ai3}, we show the dependence on $i_\mathrm{B}$ for other values of $a_\mathrm{B}$, assuming $m_3=1\,M_\mathrm{J}$. Generally, $e_\mathrm{max,A}$ is weakly dependent on $i_\mathrm{B}$ for either small $a_\mathrm{B}$ or $\mathcal{Q}_0\ll 1$ (in which case $e_\mathrm{max,A} \approx e_\mathrm{A,0}$), or large $a_\mathrm{B}$ or $\mathcal{Q}_0\gg 1$ (in which case $e_\mathrm{max,A}$ is approximately equal to the value in absence of the planet). In the intermediate regime, there is a complicated dependence on $i_\mathrm{B}$. Generally, quenching of the KL eccentricity oscillations in binary A is most efficient for close to coplanar orbits (in the case of Fig. \ref{fig:maximum_eccentricity_large_grid_test02ai3}, $0^\circ \lesssim i_\mathrm{B} \lesssim 30^\circ$ and $150^\circ \lesssim i_\mathrm{B} \lesssim 180^\circ$). Furthermore, for some values of $a_\mathrm{B}$, quenching can also be effective for $i_\mathrm{B}\approx 90^\circ$ (cf. the top middle panel of Fig. \ref{fig:maximum_eccentricity_test02} and the top left panel of Fig. \ref{fig:maximum_eccentricity_large_grid_test02ai3}). 

The dependence on $m_3$ is illustrated in the top-right panel of Fig. \ref{fig:maximum_eccentricity_test02} (for $a_\mathrm{B}=1.2\,\mathrm{AU}$ and $i_\mathrm{B}=0^\circ$). The planet mass must be large enough for shielding to be effective, which is intuitively easy to understand. In our example, a Jupiter-mass planet can effectively shield the inner binary, assuming that it is coplanar with and close to the inner binary, whereas an Earth-mass planet cannot provide shielding, regardless of its orbit.

\begin{figure}
\center
\includegraphics[scale = 0.46, trim = 8mm 0mm 0mm 0mm]{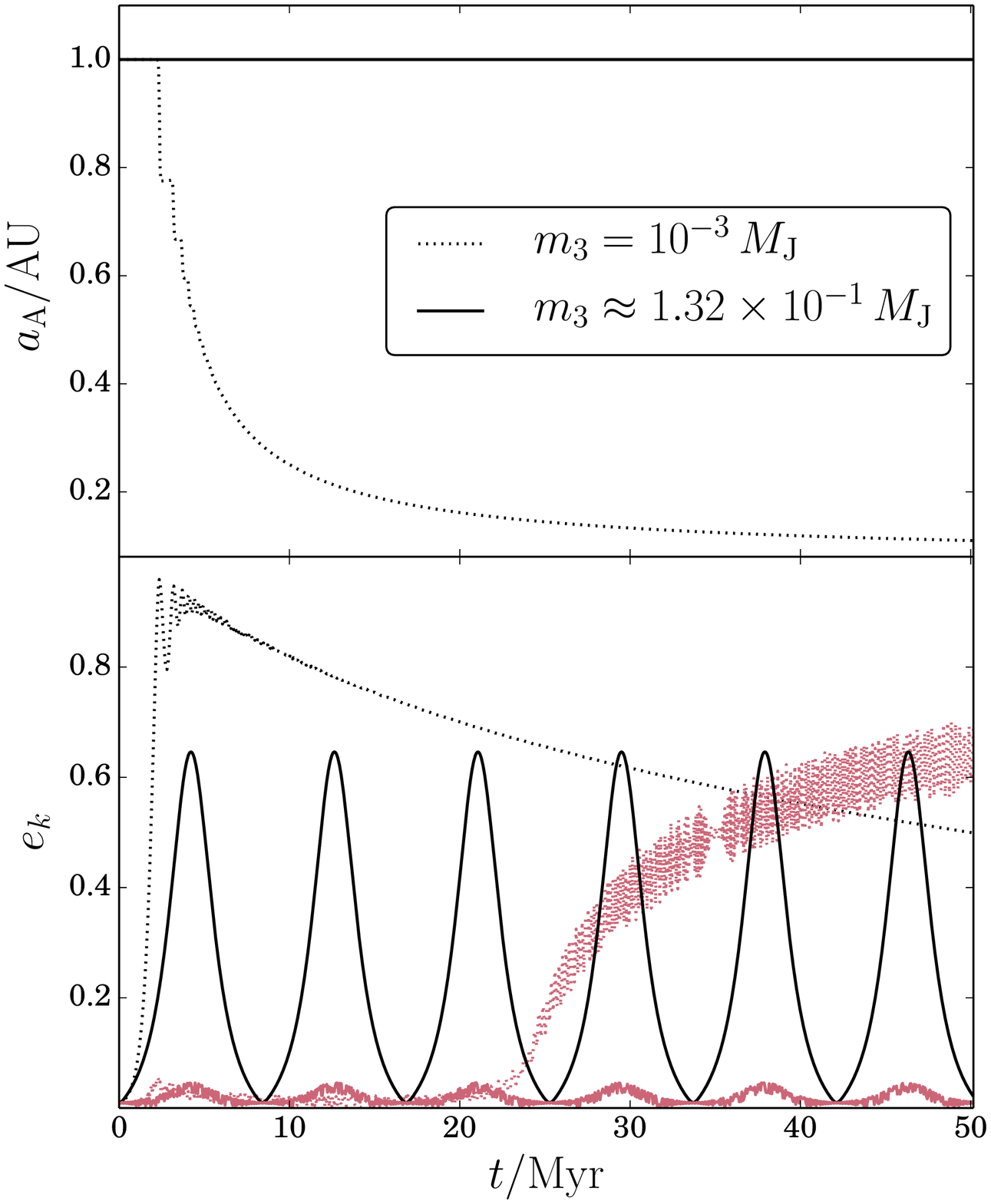}
\caption{\small Example evolution of the inner binary semimajor axis (top panel), the inner binary eccentricity (bottom panel; black lines), and the circumbinary eccentricity (bottom panel; red lines), for the triple system discussed in \S\,\ref{sect:shielding:tides}. Two planet masses are assumed: $m_3 = 10^{-3} \, M_\mathrm{J}$ (solid lines) and $m_3 \approx 1.32\times 10^{-1} \, M_\mathrm{J}$ (dashed lines). In either cases, $a_\mathrm{B} = 40 \, \mathrm{AU}$ and $i_\mathrm{B}=0^\circ$ initially.  }
\label{fig:tidal_friction_elements_time_paper_test04Lm.eps}
\end{figure}

\begin{figure*}
\center
\includegraphics[scale = 0.62, trim = 32mm 0mm 0mm 0mm]{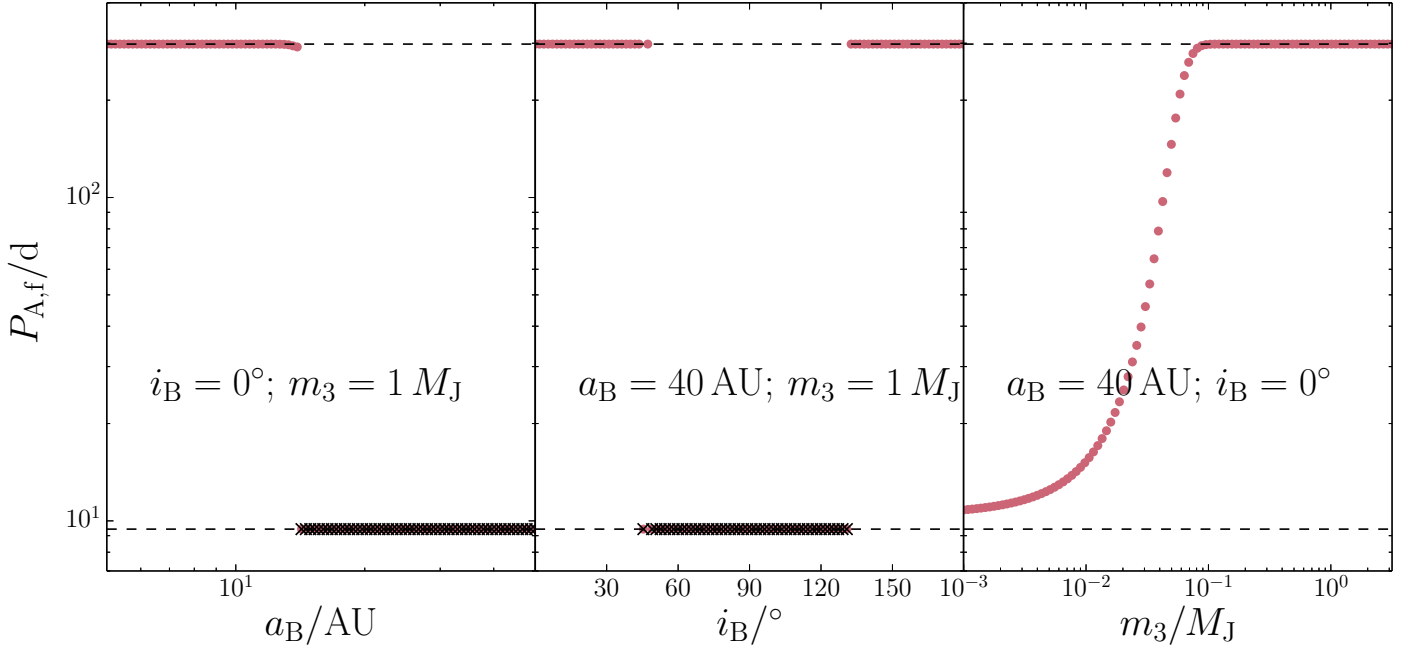}
\caption{\small The final inner orbital period for the integrations of \S\,\ref{sect:shielding:tides} (cf. the second column of Table \ref{table:init_cond1}) as a function of the initial $a_\mathrm{B}$ (left column), $i_\mathrm{B}$ (middle column) and $m_3$ (right column). In the left column $i_\mathrm{B}=0^\circ$ and $m_3=1\,M_\mathrm{J}$, in the middle column $a_\mathrm{B}=40\,\mathrm{AU}$ and $m_3=1\,M_\mathrm{J}$, and in the right column $a_\mathrm{B}=40\,\mathrm{AU}$ and $i_\mathrm{B}=0^\circ$. Cases where the planet became unstable are denoted with black crosses. }
\label{fig:tidal_friction_test04}
\end{figure*}

\subsection{A triple with a shrinking inner binary orbit}
\label{sect:shielding:tides}
Having considered in \S\,\ref{sect:shielding:kepler} the planet's ability to shield the inner binary from KL eccentricity oscillations induced by the outer orbit, we here extend the analysis by also including the effects of tidal friction in the inner binary. We choose a triple system with a relatively wide inner binary ($a_\mathrm{A}=1 \, \mathrm{AU}$ or $P_\mathrm{A} \sim 300 \, \mathrm{d}$) and tight outer binary ($a_\mathrm{C}=160\, \mathrm{AU}$), such that in the absence of a planet, the inner binary becomes highly eccentric because of KL eccentricity cycles induced by the tertiary, and shrinks to a tight binary with $P_\mathrm{A} \approx 8\, \mathrm{d}$. 

As in \S\,\ref{sect:shielding:kepler}, we carry out a set of integrations with a planet with various values of $a_\mathrm{B}$, $i_\mathrm{B}$ and $m_3$. A comprehensive list of the initial conditions is given in the second column of Table \ref{table:init_cond1}. For each system, the integration time is set to $10\,P_\mathrm{KL,AC}\approx 50 \, \mathrm{Myr}$. In contrast to \S\,\ref{sect:shielding}, here we stop the integration if the circumbinary orbit intersects with the inner binary, i.e. if $r_\mathrm{p,B} = a_\mathrm{B}(1-e_\mathrm{B}) \leq r_\mathrm{a,A} = a_\mathrm{A}(1+e_\mathrm{A})$ (neglecting the radii of the stars and the planet). We assume that in the latter case, the planet either collides with one of the stars or is ejected from the system (see also \S\,\ref{sect:discussion:unstable}), and, therefore, the subsequent evolution is equivalent to the situation without a planet, i.e. the inner binary shrinking to an inner final inner period $P_\mathrm{A,f}\approx 8 \, \mathrm{d}$.

In Fig. \ref{fig:tidal_friction_elements_time_paper_test04Lm.eps}, we show, as an example, the evolution of the inner binary semimajor axis $a_\mathrm{A}$ (top panel), the inner binary eccentricity $e_\mathrm{A}$ (bottom panel; black lines), and the circumbinary eccentricity $e_\mathrm{B}$ (bottom panel; red lines), where $a_\mathrm{B} = 40 \, \mathrm{AU}$ and $i_\mathrm{B}=0^\circ$ initially. Two planet masses are assumed: $m_3 = 10^{-3} \, M_\mathrm{J}$ (solid lines) and $m_3 \approx 1.32\times 10^{-1} \, M_\mathrm{J}$ (dashed lines). The values of $\mathcal{Q}_0$ are $\approx 55$ and $\approx 0.42$ for these values of $m_3$, respectively. 

For the low planet mass ($\mathcal{Q}_0\approx 55$), shielding is ineffective, and the inner orbit eccentricity becomes highly excited, resulting in efficient shrinkage of the orbit. Interestingly, this also increases the relative strength of the torque of the outer orbit compared to the inner orbit on the circumbinary orbit. Consequently, the eccentricity of the circumbinary orbit becomes excited at $t\sim 25 \, \mathrm{Myr}$. Although not the case here, the latter eccentricity could become high enough for the circumbinary orbit to intersect with the inner binary, and become unstable. In other words, KL cycles with tidal friction tend to destabilize the circumbinary planet, and this further reduces the chances of a circumbinary planet residing in a stable orbit around a binary that shrank due to KL cycles with tidal friction. This consequence of inner binary shrinkage on the planet orbital stability was also noted by \citet{martin_mazeh_fabrycky_15} and \citet{munoz_lai_15}. 

In contrast, for the high planet mass ($\mathcal{Q}_0\approx 0.42$), shielding is effective, and the inner binary eccentricity does not become high enough for efficient tidal friction. Nevertheless, $e_\mathrm{A}$ still oscillates with an amplitude of $\approx 0.6$. The circumbinary orbit eccentricity does not become excited in this case.

In the left, middle and right panels of Fig. \ref{fig:tidal_friction_test04}, we show the final inner period $P_\mathrm{A,f}$ as a function of $a_\mathrm{B}$ (with $i_\mathrm{B}=0^\circ$ and $m_3=1\,M_\mathrm{J}$), $i_\mathrm{B}$ (with $a_\mathrm{B}=40\,\mathrm{AU}$ and $m_3=1\,M_\mathrm{J}$), and $m_3$ (with $a_\mathrm{B}=40\,\mathrm{AU}$ and $i_\mathrm{B}=0^\circ$), respectively. Cases where the planet became unstable are denoted with black crosses. There is a strong dependence of the final inner period on the circumbinary parameters. If $a_\mathrm{B}$ is sufficiently small (the regime $\mathcal{Q}_0\ll1$), shielding is effective, and the inner binary does not shrink. For larger $a_\mathrm{B}$, the planet becomes unstable, and is therefore unable to shield the inner binary. Note that in this regime of $\mathcal{Q}_0\gtrsim 1$, shielding would have been ineffective even if the circumbinary orbit would be stable (cf. Fig. \ref{fig:maximum_eccentricity_test02}).

Similarly, there is a strong dependence on $i_\mathrm{B}$. For coplanar (either prograde or retrograde) circumbinary orbits, shielding can be effective (if $a_\mathrm{B}$ is small enough) and the inner binary does not shrink. However, for highly inclined orbits, the planetary orbit is unstable and, subsequently, the inner binary shrinks. The boundaries between these regimes, $i_\mathrm{B} \approx 45^\circ$ and $i_\mathrm{B} \approx 135^\circ$, are very similar to the critical values for eccentricity excitation due to the KL mechanisms (assuming quadrupole order). We note that, for much more compact circumbinary planets with a small ratio $a_\mathrm{B}/a_\mathrm{A}$, instability can occur for even coplanar systems \citep{li_ea_14}. 

For the value of $a_\mathrm{B}$ and $i_\mathrm{B}$ in the right panel of Fig. \ref{fig:tidal_friction_test04}, the planetary orbit is stable for all assumed masses $m_3$. Effective shielding, and hence shrinkage of the inner orbit, can only occur if the planet is sufficiently massive, in this case, if $m_3 \gtrsim 0.1 \, M_\mathrm{J}$. 

\begin{figure}
\center
\includegraphics[scale = 0.48, trim = 10mm 0mm 0mm 0mm]{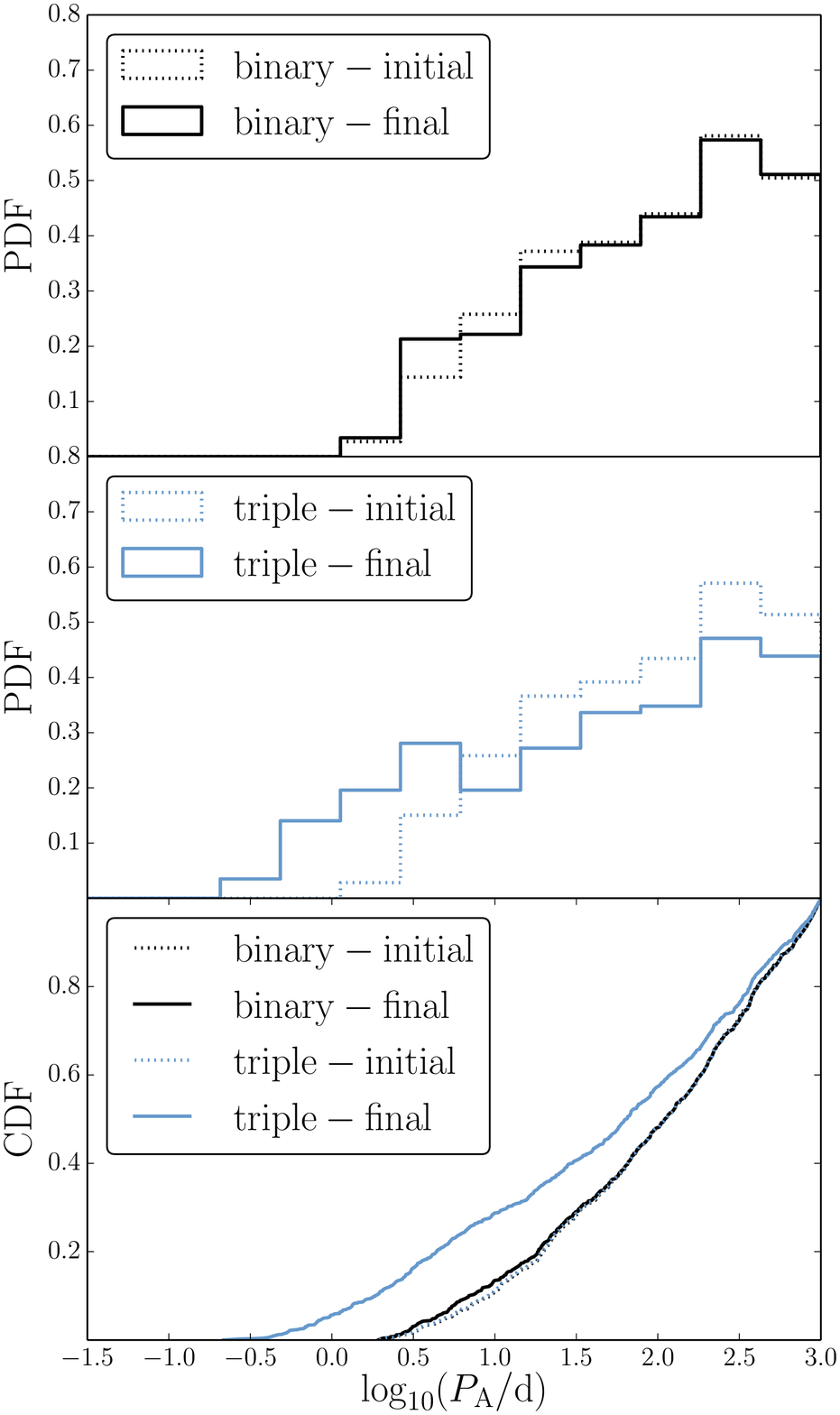}
\caption{\small The initial (dotted lines) and final (solid lines) inner period distributions for the sampled triples as described in \S\,\ref{sect:pop_syn:init_cond}. Top row (black lines): isolated binary evolution, i.e. taking into account only tidal friction. Middle row (blue lines): isolated triple evolution, i.e. also taking into account the torque of the outer orbit. Cumulative distributions for both cases are shown in the bottom panel. }
\label{fig:period_distributions_test02_mode_B_and_T}
\end{figure}

\begin{figure}
\center
\includegraphics[scale = 0.42, trim = 10mm 0mm 0mm 0mm]{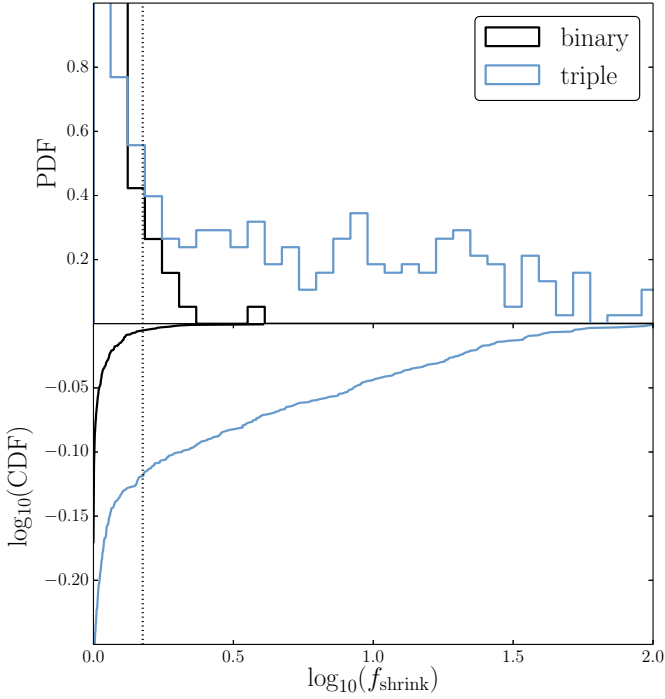}
\caption{\small The factor $f_\mathrm{shrink} \equiv a_\mathrm{A,i}/a_\mathrm{A,f}$ with which the inner binary semimajor axis shrinks for the sampled triples as described in \S\,\ref{sect:pop_syn:init_cond}. Black (blue) lines correspond to isolated binary (triple) evolution. The black vertical dotted line indicates the value of $f_\mathrm{shrink}$ chosen for the selection of systems with substantially shrinking inner binaries. }
\label{fig:f_shrink_distributions_test02_mode_B_and_T}
\end{figure}

\begin{figure}
\center
\includegraphics[scale = 0.48, trim = 10mm 0mm 0mm 0mm]{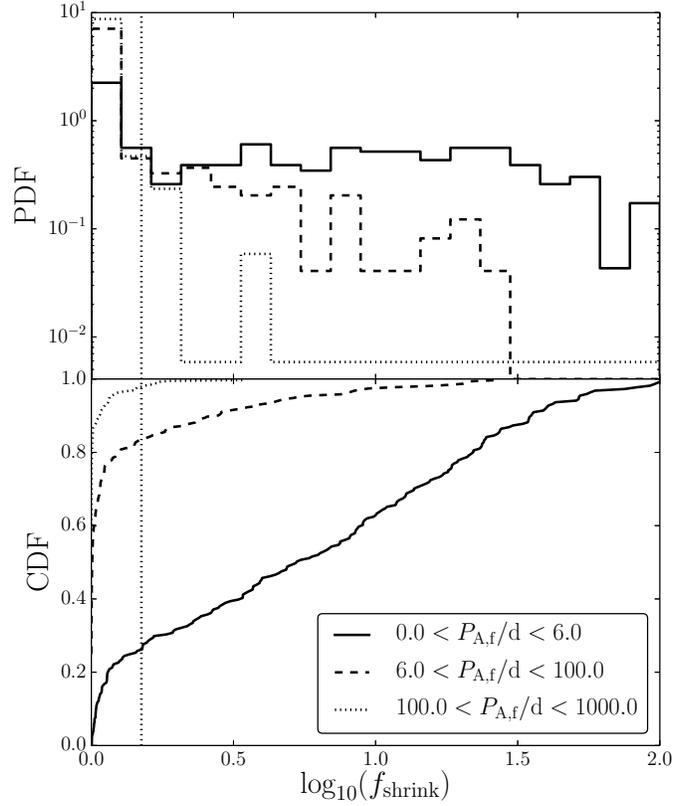}
\caption{\small The factor $f_\mathrm{shrink} \equiv a_\mathrm{A,i}/a_\mathrm{A,f}$ with which the inner binary semimajor axis shrinks for the sampled triples as described in \S\,\ref{sect:pop_syn:init_cond}, for isolated triple evolution as in Fig. \ref{fig:f_shrink_distributions_test02_mode_B_and_T}, and here binned with respect to the final inner orbital period $P_\mathrm{A,f}$. Three bins in $P_\mathrm{A,f}$ are assumed, indicated in the legend in the bottom panel. The black vertical dotted line indicates the value of $f_\mathrm{shrink}$ chosen for the selection of systems with substantially shrinking inner binaries. }
\label{fig:f_shrink_P_A_final_test02_mode_B_and_T.eps}
\end{figure}

\begin{figure}
\center
\includegraphics[scale = 0.48, trim = 10mm 0mm 0mm 0mm]{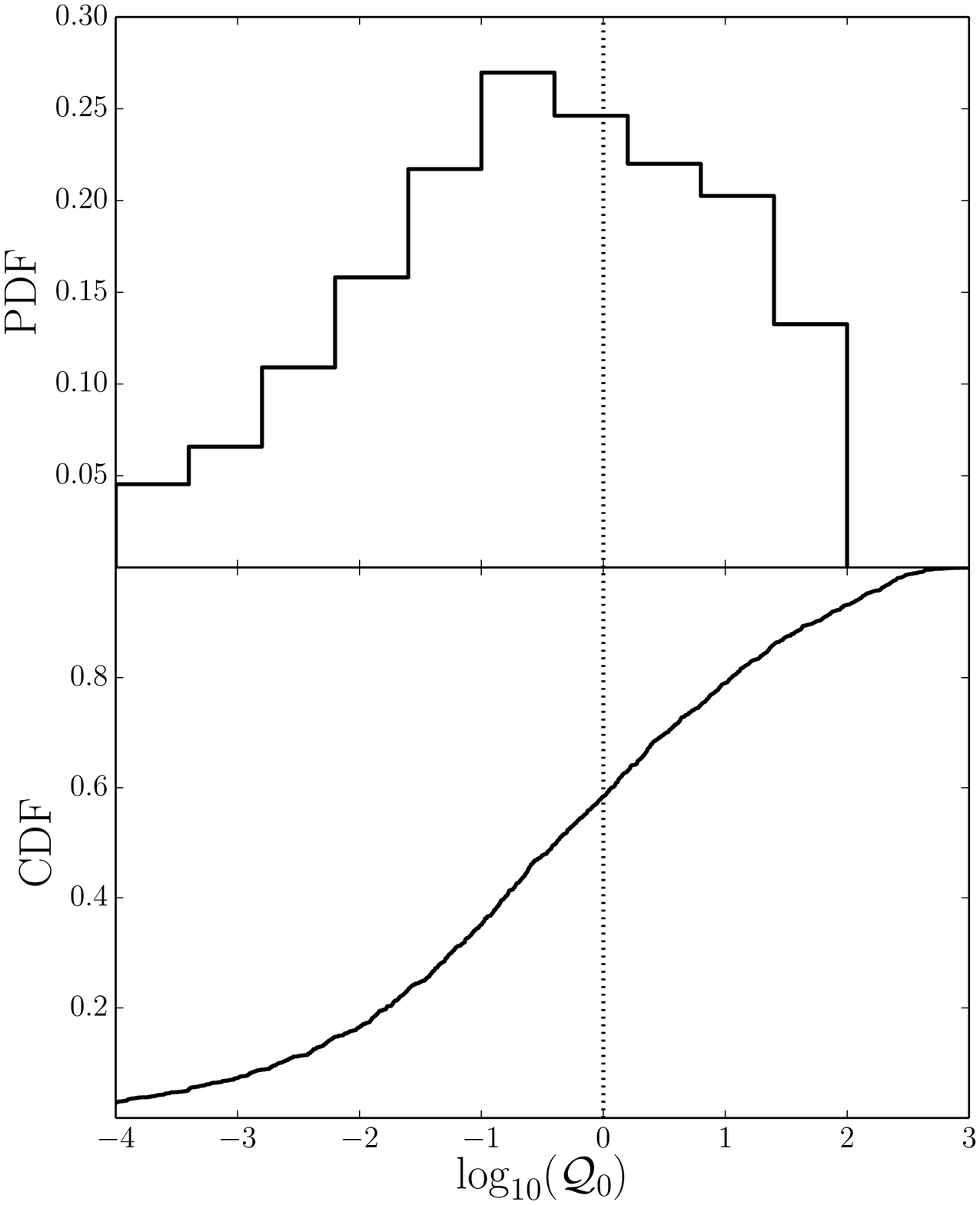}
\caption{\small The distribution of $\mathcal{Q}_0$ (cf. equation~\ref{eq:Q0}) for the systems that exceeded the imposed maximum wall time limit of 15 min. The black vertical dotted line shows the value $\mathcal{Q}_0=1$ for which, for these systems, planets were assumed to be fully effective ($\mathcal{Q}_0<1$) or ineffective ($\mathcal{Q}_0>1$) at shielding. }
\label{fig:grid_dependence_R0_flag_100_paper_test02_mode_Q}
\end{figure}

\section{Planets in triples with short-period inner binaries}
\label{sect:pop_syn}

\subsection{Initial conditions}
\label{sect:pop_syn:init_cond}
In \S\,\ref{sect:shielding}, we demonstrated the planet's ability to shield the inner binary assuming a fixed triple system, and used specific worked-out examples to understand the evolution of circumbinary planets affected by a third companion. In the following, we study the effect of circumbinary planets in the more general case, and consider a large orbital phase-space through the study of a population of triples, for which the inner binary, in absence of a planet, would be affected by KL cycles with tidal friction. 

For stellar triples, a number of surveys of solar-type MS stars \citep{duquennoy_mayor_91,tokovinin_ea_06,raghavan_ea_10,tokovinin_14a} have given insight into the initial distributions of the masses and orbital parameters. However, initial distributions for a circumbinary planet in orbit of the inner binary, for which observations are currently strongly biased, are very poorly constrained. Therefore, we generate initial conditions using a combination of Monte Carlo sampling for the stellar triple system, and fixed grids for the circumbinary planet. Our procedure consists of the following four steps (a succinct summary is given in the fourth column of Table \ref{table:init_cond1}).

1. A stellar triple system is sampled, similarly to \citet{fabrycky_tremaine_07,naoz_fabrycky_14}. For the inner binary, the primary mass $m_1$ is sampled between 0.5 and 1.2 $\mathrm{M}_\odot$ assuming a Salpeter distribution, $\mathrm{d}N/\mathrm{d}m_1\propto m^{-2.35}$ \citep{salpeter_58}. The secondary mass $m_2$ is sampled from $m_2=q_\mathrm{in} m_1$ with the constraint $0.1<m_2/\mathrm{M}_\odot<1.2$, where $q_\mathrm{in}$ is sampled from a flat distribution with $0<q_\mathrm{in}<1$. The mass of the tertiary star, $m_4$, with $0.1<m_4/\mathrm{M}_\odot<1.2$, is sampled from $m_4 = (m_1+m_2)q_\mathrm{out}$, where $q_\mathrm{out}$ is sampled from a flat distribution $0<q_\mathrm{out}<1$. The stellar radii, $R_1$, $R_2$ and $R_4$, are computed initially, and at subsequent times, using the stellar evolution code \textsc{SeBa} as described in \S\,\ref{sect:methods}. The spins $\boldsymbol{\Omega}_1$ and $\boldsymbol{\Omega}_2$ of the stars in the inner binary are initially assumed to be parallel with the inner orbital angular momentum, and the spin periods are assumed to be $2\pi/||\boldsymbol{\Omega}_k|| = 10 \, \mathrm{d}$. The tidal quality factors $Q'_k$ for the inner binary primary and secondary stars (cf. \S\,\ref{sect:methods}) are sampled linearly between $5.5\times10^5$ and $2\times 10^6$, adopted from \citealt{ogilvie_lin_07}, who constrained $Q'_k$ using observations of spectroscopic binaries by \citet{meibom_mathieu_05}. The gyration radii $r_{\mathrm{g},k}$ for the primary and secondary stars are assumed to be $r_{\mathrm{g},k}=0.08$.

The inner and outer orbital periods, $P_\mathrm{A}$ and $P_\mathrm{C}$, are both sampled from a lognormal distribution with mean $\log_{10}(P_k/\mathrm{d}) = 5.03$, standard deviation $\sigma_{\log_{10}(P_k/\mathrm{d})} = 2.28$ and range $-2 < \log_{10} (P_k/\mathrm{d}) < 10$. The latter distribution is consistent with the orbital periods of binaries as found by \citet{duquennoy_mayor_91,raghavan_ea_10,tokovinin_14a}. We choose to impose the further restriction of $P_\mathrm{A}<10^3\,\mathrm{d}$, because most of the progenitor systems of short-period binaries produced through KL cycles with tidal friction initially have $P_\mathrm{A}<10^3\, \mathrm{d}$ \citep{fabrycky_tremaine_07}. The corresponding semimajor axes, $a_\mathrm{A}$ and $a_\mathrm{C}$, are computed from the orbital periods using Kepler's law (neglecting the planet mass $m_3$). 

The eccentricities $e_\mathrm{A}$ and $e_\mathrm{C}$ are sampled from a Rayleigh distribution, $\mathrm{d} N/\mathrm{d} e_k \propto e_k \exp(-\beta e_k^2)$, with rms $\langle e_k^2 \rangle^{1/2} = \beta^{-1/2} = 0.33$, consistent with the results of \citealt{raghavan_ea_10}. In \citet{raghavan_ea_10}, in contrast with \citet{duquennoy_mayor_91}, the eccentricity distribution is found not to be substantially different for periods $P_k< 10^3 \, \mathrm{d}$ and periods $P_k> 10^3 \, \mathrm{d}$. Therefore, we choose not to sample eccentricities from a thermal distribution for binaries with $P_k>10^3 \, \mathrm{d}$ as in \citet{fabrycky_tremaine_07}. 

The inner and outer orbits are assumed to be randomly oriented, i.e. their arguments of pericentre $\omega_k$ and $\Omega_k$ are sampled from a random distribution, and their mutual inclination $i_\mathrm{AC}$ is sampled from a distribution that is linear in $\cos(i_\mathrm{AC})$. Note that, without loss of generality, we fix the inner orbit to be aligned with the $z$ axis of the coordinate system, i.e. we set $i_\mathrm{A}=0^\circ$. The {\it initial} mutual inclinations $i_\mathrm{AB}$ and $i_\mathrm{AC}$ are therefore equal to the initial individual inclinations $i_\mathrm{B}$ and $i_\mathrm{C}$, respectively. 

Sampled triple systems are rejected in the following cases at this stage.
\begin{itemize}
\item The stability criterion of \citealt{mardling_aarseth_01} is not satisfied.
\item In isolation, the stars in the inner binary would experience strong tidal interaction, i.e. $a_\mathrm{A}(1-e_\mathrm{A}) < 3 \, (R_1+R_2)$.
\item In isolation, the stars in the inner binary would fill their Roche lobe, i.e. $R_k < R_{\mathrm{L},k}$ for $k\in\{1,2\}$, where $R_{\mathrm{L},k}$ is the Roche lobe radius computed at pericentre according to the analytic fitting formulae given by \citet{sepinsky_ea_07}. 
\end{itemize}

2. For the sampled systems of step (1), additional selection is made based on the outcome of the integration of isolated systems. This is to ensure that in the absence of the planet, the system would remain dynamically stable, and that the inner binary is shrunk substantially because of KL cycles with tidal friction. The integration time for each system is sampled linearly between 1 and 10 Gyr. In addition, the following stopping conditions are always imposed.
\begin{itemize}
\item The stars in the inner binary collide or fill their Roche lobe, computed as above.
\item One of the three stars evolves past the MS. 
\end{itemize}

\begin{figure*}
\center
\includegraphics[scale = 0.68, trim = 20mm 0mm 0mm 0mm]{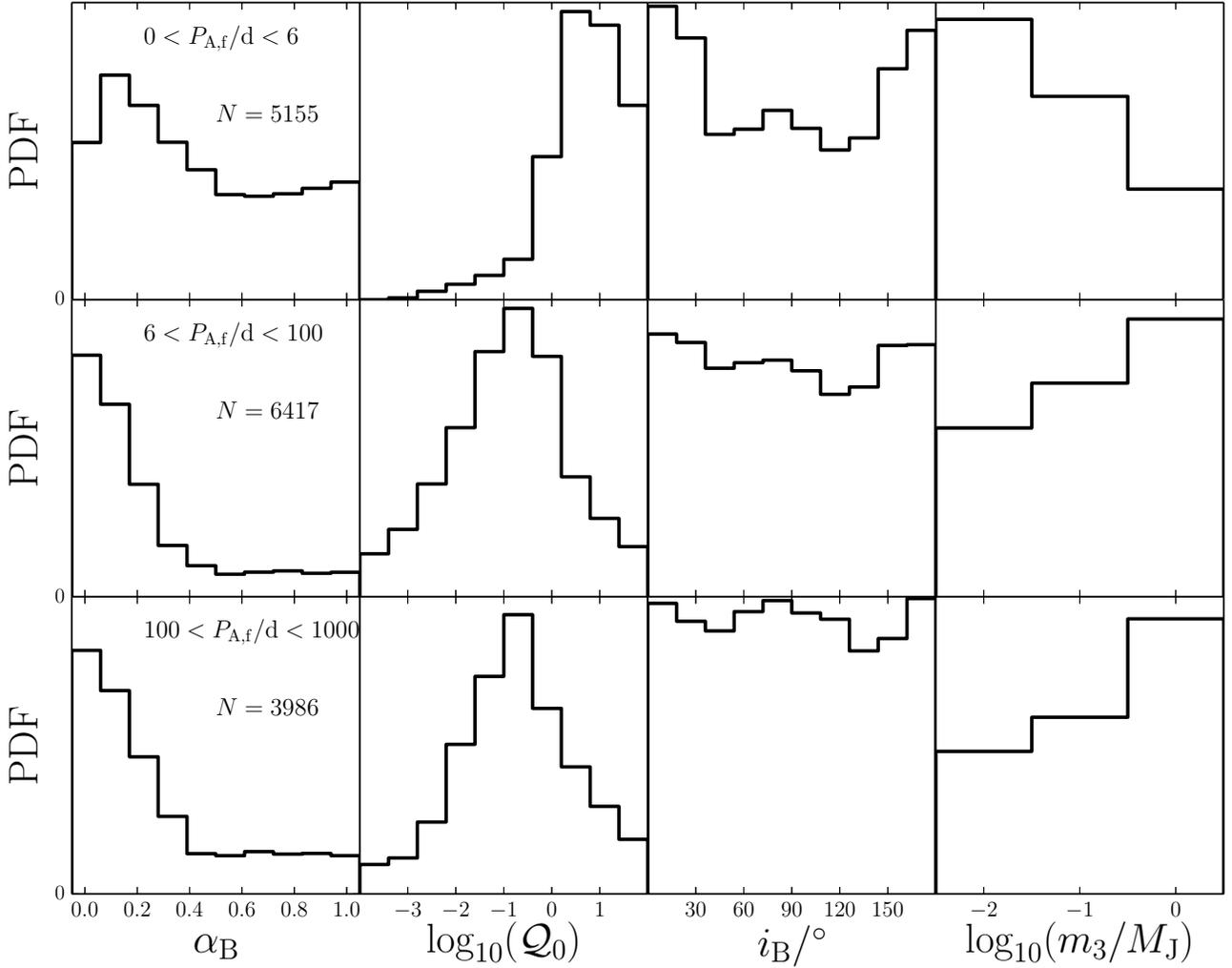}
\caption{\small The distributions of the parameters $\alpha_\mathrm{B}$ (first column), $\mathcal{Q}_0$ (second column), $i_\mathrm{B}$ (third column) and $m_3$ (fourth column) for systems in which the orbit of the planet remains stable during the entire evolution, and no other stopping conditions occurred. Systems are binned with respect to the final inner orbital period $P_\mathrm{A,f}$; each row corresponds to a different bin, indicated in the first column. Also indicated in the first column is the number of systems in the corresponding bin of $P_\mathrm{A,f}$. }
\label{fig:grid_dependence_P_A_paper_test02_mode_Q}
\end{figure*}

First, the sampled systems are integrated taking into account only the inner binary evolution, i.e. stellar and tidal evolution. From these systems, we reject those for which the stars in the inner binary collide, fill their Roche lobe, or evolve past the MS. The former two cases can occur because of tidal friction; note that our inner binaries are initially not circular, nor synchronized [see \S\,\ref{sect:discussion:other_dis_MB} and Appendix \ref{app:MB} for discussion on magnetic braking (MB) effects].

Secondly, the sampled systems are integrated taking into account the evolution of the triple system without a circumbinary planet, i.e. taking into account KL cycles in the inner and outer binaries, and tidal evolution in the inner binary. From these systems, we also reject those for which the inner binary stars collide or fill their Roche Lobe. 

In Fig. \ref{fig:period_distributions_test02_mode_B_and_T}, we show the initial (dotted lines) and final (solid lines) inner period distributions for the case of an isolated inner binary system (black lines), and an isolated triple system (blue lines). Consistent with previous studies on KLCTF \citep{mazeh_shaham_79,eggleton_ea_01,fabrycky_tremaine_07,naoz_fabrycky_14}, the effect of the tertiary star is to produce an enhancement of systems with periods roughly between 1 and 6 d. 

3. We select systems for which, in the case of an isolated triple, the inner binary shrinks substantially during the evolution. To quantify this criterion, we show in Fig. \ref{fig:f_shrink_distributions_test02_mode_B_and_T} the distribution of the factor $f_\mathrm{shrink}\equiv a_\mathrm{A,i}/a_\mathrm{A,f}$ with which the inner binary semimajor axis decreases during the evolution for the case of isolated triples. Based on this distribution, we select systems for which $f_\mathrm{shrink} \geq 1.5$. Effectively, this implies that the majority of the systems evolve to a system with a short-period inner period, $1<P_\mathrm{A,f}<6$. This is demonstrated in Fig. \ref{fig:f_shrink_P_A_final_test02_mode_B_and_T.eps}, where we show the distribution of $f_\mathrm{shrink}$ in the triple case, for different bins of the final inner orbital period. 

4. For each of the $N_\mathrm{tr}=192$ remaining sampled triple systems, we define a three-dimensional grid in $(a_\mathrm{B},i_\mathrm{B},m_3)$, with size $10\times10\times3=300$. The inclinations $i_\mathrm{B}$ range from 0 to $180^\circ$, and for the planet mass $m_3$ we assume three values, $10^{-2}$, $10^{-1}$ and $10^0 \, M_\mathrm{J}$. For $a_\mathrm{B}$, we impose the lower and upper limits $a_\mathrm{B,l}$ and $a_\mathrm{B,u}$, respectively. 

The lower limit is assumed to be $a_\mathrm{B,l} = 1.5 \, a_\mathrm{B,crit,AB}$, where $a_\mathrm{B,crit,AB}$ is the critical semimajor axis for dynamical stability of the planet in a coplanar orbit around the inner binary, and which is adopted from the fitting formula given by \citep{holman_wiegert_99}. The latter formula is a function of the inner binary mass ratio $\mu = m_2/m_1$ and eccentricity $e_\mathrm{A}$. The factor of 1.5 in $a_\mathrm{B,l} = 1.5 \, a_\mathrm{B,crit,AB}$ is a `safety factor'; close to the regime of dynamical stability, the orbits are non-Keplerian (e.g. \citealt{lee_peale_06,leung_lee_13,bromley_kenyon_15}), and, therefore, the orbit-averaged approach no longer applies. The upper limit is assumed to be $a_\mathrm{B,u} = 0.9\,a_\mathrm{B,crit,BC}$, where $a_\mathrm{B,crit,BC}$ is the largest possible value of $a_\mathrm{B}$ for dynamical stability with respect to the orbit of the tertiary star. The latter value is estimated by applying the \citealt{mardling_aarseth_01} criterion to the BC pair, with an `outer' mass ratio of $q_\mathrm{out} = m_4/(m_1+m_2+m_3)$.

In the analysis below, we use the dimensionless quantity $\alpha_\mathrm{B}$ which is closely related to $a_\mathrm{B}$, and is defined as
\begin{align}
\alpha_\mathrm{B} \equiv \frac{a_\mathrm{B} - a_\mathrm{B,l}}{a_\mathrm{B,u} - a_\mathrm{B,l}}.
\label{eq:alpha_B}
\end{align}
This quantity is motivated by the fact that the allowed range of initial $a_\mathrm{B}$ for dynamical stability varies per triple system. By construction, this is not the case for $\alpha_\mathrm{B}$, which, for any sampled triple, ranges between $\alpha_\mathrm{B}=0$ ($a_\mathrm{B} = 1.5 \,a_\mathrm{B,crit,AB}$, i.e. the planet's orbit is close to the dynamical stability limit with respect to the inner binary), and $\alpha_\mathrm{B}=1$ ($a_\mathrm{B} = 0.9 \,a_\mathrm{B,crit,BC}$, i.e. the planet's orbit is close to the dynamical stability limit with respect to the outer binary). In other words, by using $\alpha_\mathrm{B}$, we can evaluate the effect of the circumbinary planet for a population of triples with different parameters.

For the $N_\mathrm{quad} = 300\times 192 = 57600$ integrations with a planet, the following additional stopping conditions are set.
\begin{itemize}
\item The circumbinary orbit is unstable with respect to the inner and outer binaries according to the \citep{mardling_aarseth_01} criterion. 
\item The circumbinary orbit intersects with the inner binary, i.e. $a_\mathrm{B}(1-e_\mathrm{B}) \leq a_\mathrm{A}(1+e_\mathrm{A})$, and is therefore unlikely to be stable. 
\end{itemize}

For some quadruple systems, the integration proceeded slowly. We therefore imposed a maximum wall time of 15 min per system, which was reached for $19\%$ of the integrated systems. In Fig. \ref{fig:grid_dependence_R0_flag_100_paper_test02_mode_Q}, we show the distribution of $\mathcal{Q}_0$ (cf. equation~\ref{eq:Q0}) for the systems in which the maximum wall time was exceeded. For $\approx 60\%$ of these systems, $\mathcal{Q}_0 \lesssim 1$. As shown previously in \S\,\ref{sect:shielding}, for $\mathcal{Q}_0 \lesssim 1$, shielding is typically effective, whereas it is not for $\mathcal{Q}_0 \gtrsim 1$. Therefore, for the purposes of the analysis below, we will assume that in the systems that exceeded the integration wall time, if $\mathcal{Q}_0\lesssim 1$, shielding is completely effective and hence the inner orbital period does not change (i.e. $P_\mathrm{A,f} = P_\mathrm{A,i}$), and if $\mathcal{Q}_0\gtrsim 1$, shielding is completely ineffective and hence the inner orbital period shrinks as much as would be the case without a planet. Clearly, this is approximate because in reality, there is no sharp transition between the two regimes and there is also a dependence on the inclination $i_\mathrm{B}$ (cf. Fig. \ref{fig:maximum_eccentricity_test02}). However, we do not expect that the approximation strongly affects our conclusions. 

\subsection{Results}
\label{sect:pop_syn:results}
\subsubsection{Initial parameters for surviving planets as a function of the final binary period}
\label{sect:pop_syn:results:planets_short_period}
In Fig. \ref{fig:grid_dependence_P_A_paper_test02_mode_Q}, we show the distributions of the initial parameters $\alpha_\mathrm{B}$ (cf. equation~\ref{eq:alpha_B}), $\mathcal{Q}_0$ (cf. equation~\ref{eq:Q0}), $i_\mathrm{B}$ and $m_3$ in the first through fourth columns, respectively, for systems in which the orbit of the planet remains stable during the entire evolution according to the criteria discussed in \S\,\ref{sect:pop_syn:init_cond}. Each row in Fig. \ref{fig:grid_dependence_P_A_paper_test02_mode_Q} corresponds to a different bin of the final inner orbital period, $P_\mathrm{A,f}$. \\

For short periods, the planet tends to be further away from the inner binary (i.e. $\alpha_\mathrm{B}$ and $\mathcal{Q}_0$ are typically larger) compared to longer periods. This trend is compatible with the lack of tight coplanar planets around short-period binaries, although for our sampled population, there is certainly no {\it absolute} lack of planets in tight (i.e. $\alpha_\mathrm{B}\lesssim 0.3$) orbits around short-period binaries. 

In fact, for $1\, \mathrm{d} < P_\mathrm{A,f}<6\, \mathrm{d}$, the distribution of $\alpha_\mathrm{B}$ is still peaked around $\alpha_\mathrm{B}\approx 0.1$. There are two important factors that contribute to this.
\begin{enumerate}
\item In our initial triple population, there is a non-negligible number of systems with {\it initial} inner periods $1\, \mathrm{d} < P_\mathrm{A,i}<6\, \mathrm{d}$ (cf. Fig. \ref{fig:period_distributions_test02_mode_B_and_T}). Evidently, for these systems, a possible outcome is a final period $1 \, \mathrm{d} <P_\mathrm{A,f}<6\, \mathrm{d}$ when the inner orbit did not shrink due to effective shielding by the planet, likely corresponding to a massive planet in a tight and/or coplanar orbit. In other words, in principle, a short-period binary can simply be born with a tight and coplanar planet, and the planet would prevent the inner orbital period from becoming even shorter. Given the observational evidence for a third stellar companion around most short-period binaries \citep{tokovinin_ea_06}, one might consider the possibility that such binaries can not primordially form without the effects of KLCTF. If this is the case, then the primordially short-period binaries in our simulations should not be considered, and therefore, we would expect no tight coplanar circumbinary planets to exist at all around short-period binaries.
\item For $\alpha_\mathrm{B} \gtrsim 0.3$ or $\mathcal{Q}_0 \gtrsim 10^0$, the planetary orbit for a large fraction of systems becomes highly eccentric and intersects with the inner binary, and likely becomes unstable (cf. Sections\,\ref{sect:pop_syn:results:unstable} and \ref{sect:discussion:unstable}). These cases are not included in Fig. \ref{fig:grid_dependence_P_A_paper_test02_mode_Q}, and this produces a bias for an {\it absence} of stable systems with $\alpha_\mathrm{B} \gtrsim 0.3$. 
\end{enumerate}

To take these complicating factors into account, we introduce the `shielding efficiency' $\eta_\mathrm{shield}$ which quantifies the planet's ability to shield the inner binary from KL eccentricity oscillations induced by the tertiary star, defined as
\begin{align}
\label{eq:eta_shield}
\eta_\mathrm{shield} \equiv \frac{f_\mathrm{shrink,triple} - f_\mathrm{shrink,quad}}{f_\mathrm{shrink,triple} - 1}.
\end{align}
Here, $f_\mathrm{shrink,triple}$ is the factor with which the inner binary semimajor axis decreases in the absence of a planet, and $f_\mathrm{shrink,quad}$ is the corresponding factor with a planet present.  Note that we restrict to systems for which $f_\mathrm{shrink,triple} \geq 1.5$. If the planet is able to fully shield the inner binary, $f_\mathrm{shrink,quad} = 1$, and $\eta_\mathrm{shield} = 1$. On the other hand, when shielding is completely ineffective, $f_\mathrm{shrink,quad} = f_\mathrm{shrink,triple}$, and $\eta_\mathrm{shield} = 0$. For some systems, we find that the planet can {\it enhance} the inner binary eccentricy excitations; in this case, $\eta_\mathrm{shield}<0$ (cf. \S\,\ref{sect:pop_syn:results:reversed_shielding}). 

\begin{figure*}
\center
\includegraphics[scale = 0.68, trim = 20mm 0mm 0mm 0mm]{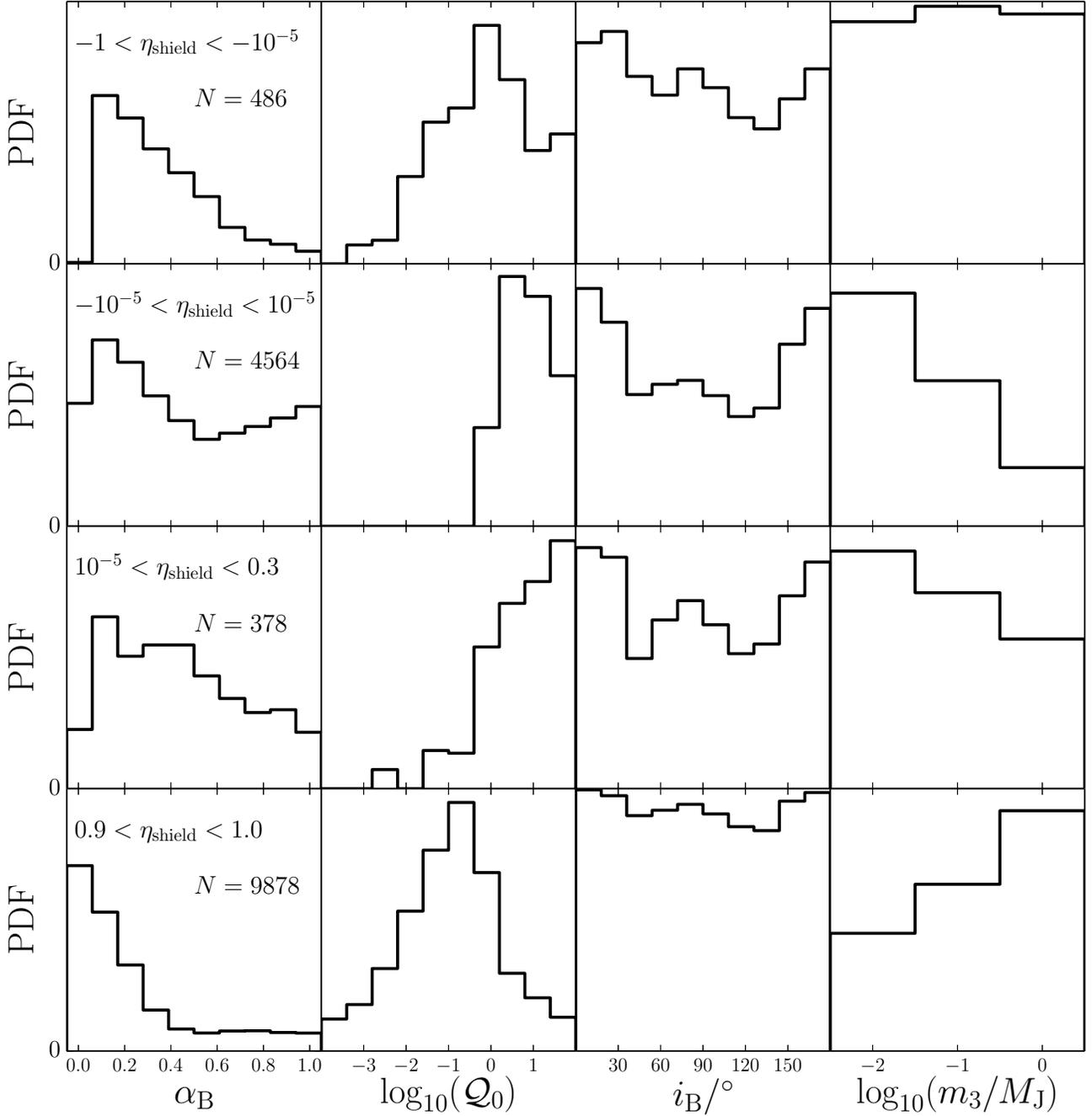}
\caption{\small The distributions of the parameters $\alpha_\mathrm{B}$ (first column), $\mathcal{Q}_0$ (second column), $i_\mathrm{B}$ (third column) and $m_3$ (fourth column), where systems are binned with respect to the shielding efficiency $\eta_\mathrm{shield}$ (cf. equation~\ref{eq:eta_shield}); each row corresponds to a different bin, indicated in the first column. Also indicated in the first column is the number of systems in the corresponding bin of $\eta_\mathrm{shield}$. }
\label{fig:grid_dependence_eta_shield_paper_test02_mode_Q}
\end{figure*}

In Fig. \ref{fig:grid_dependence_eta_shield_paper_test02_mode_Q}, we show the distributions of $\alpha_\mathrm{B}$, $\mathcal{Q}_0$, $i_\mathrm{B}$ and $m_3$, where the systems are now binned with respect to $\eta_\mathrm{shield}$. When shielding is ineffective ($\eta_\mathrm{shield} \approx 0$, cf. the second row in Fig. \ref{fig:grid_dependence_eta_shield_paper_test02_mode_Q}), $\mathcal{Q}_0$ is invariably $\gtrsim 1$ and $m_3$ tends to be low (i.e., the large value of $m_3 = 1\, M_\mathrm{J}$ is disfavoured). In contrast, when shielding is highly effective ($\eta_\mathrm{shield} \approx 1$, cf. the last row in Fig. \ref{fig:grid_dependence_eta_shield_paper_test02_mode_Q}), $\mathcal{Q}_0$ is typically $\lesssim 1$ (with a peak near $\mathcal{Q}_0 = 0.1$), and $m_3$ tends to be high (i.e., the large value of $m_3 = 1\, M_\mathrm{J}$ is favoured). 

\begin{figure}
\center
\includegraphics[scale = 0.45, trim = 10mm 0mm 0mm 0mm]{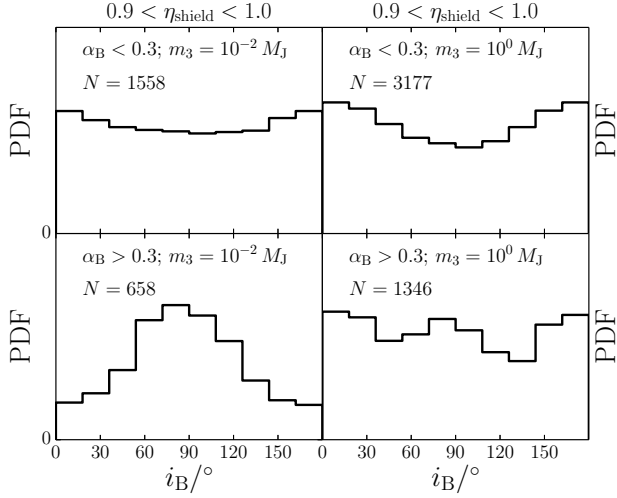}
\caption{\small The distributions of the initial $i_\mathrm{B}$ for $0.9<\eta_\mathrm{shield}<1$ (cf. Fig. \ref{fig:grid_dependence_eta_shield_paper_test02_mode_Q}), distinguishing between different ranges of $\alpha_\mathrm{B}$ and $m_3$. In the top (bottom) row, $\alpha_\mathrm{B}<0.3$ ($\alpha_\mathrm{B}>0.3$); in the left (right) column, $m_3 = 10^{-2} M_\mathrm{J}$ ($m_3 = 10^{0} M_\mathrm{J}$). The number of systems in each case is indicated in each panel. }
\label{fig:grid_dependence_eta_shield_paper_subgroups_bins_test02_mode_Q}
\end{figure}

The distribution of the initial $i_\mathrm{B}$ in Fig. \ref{fig:grid_dependence_eta_shield_paper_test02_mode_Q} is a strong function of $\alpha_\mathrm{B}$ and $m_3$. In Fig. \ref{fig:grid_dependence_eta_shield_paper_subgroups_bins_test02_mode_Q}, we show, for $0.9<\eta_\mathrm{shield}<1$, the inclination distribution for different ranges of $\alpha_\mathrm{B}$ and $m_3$. For $\alpha_\mathrm{B} < 0.3$, there is a tendency for coplanar orbits, for both low-mass and massive planets. For planets in wider orbits, $\alpha_\mathrm{B} > 0.3$, the distribution of $i_\mathrm{B}$ is strongly peaked towards $i_\mathrm{B} \approx 90^\circ$ for low planet masses, whereas for high planet masses, the distribution of $i_\mathrm{B}$ is much less peaked. These trends are qualitatively consistent with the smaller number of integrations that were carried out in \S\,\ref{sect:shielding} for fixed parameters of the stellar triple. 

\begin{figure*}
\center
\includegraphics[scale = 0.65, trim = 15mm 0mm 0mm 0mm]{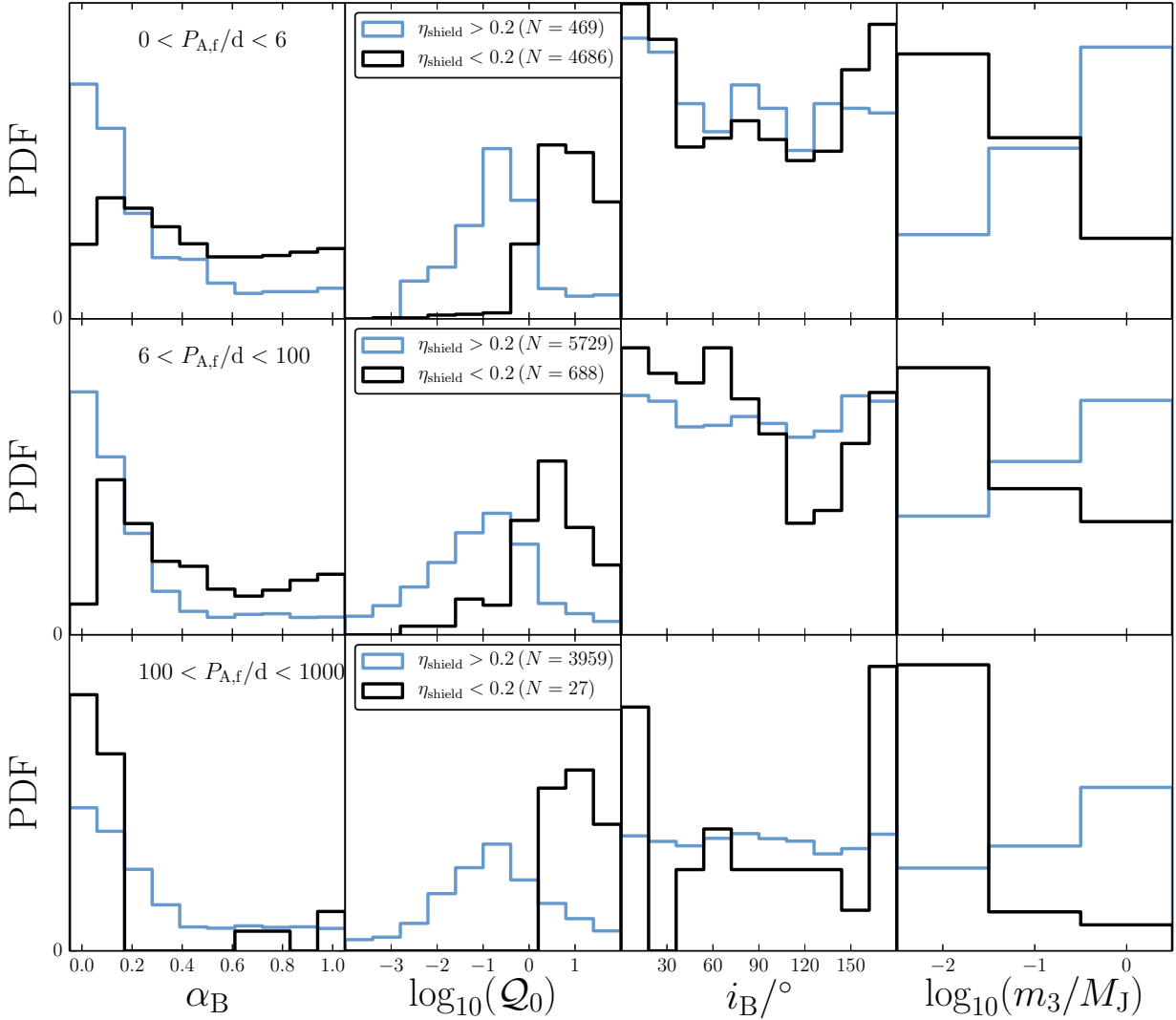}
\caption{\small The distributions of the parameters $\alpha_\mathrm{B}$ (first column), $\mathcal{Q}_0$ (second column), $i_\mathrm{B}$ (third column) and $m_3$ (fourth column) for systems in which the orbit of the planet remains stable during the entire evolution, and no other stopping conditions occurred. Systems are binned with respect to the final inner orbital period $P_\mathrm{A,f}$ as in Fig. \ref{fig:grid_dependence_P_A_paper_test02_mode_Q}; here, we also make a distinction between a low shielding efficiency ($\eta_\mathrm{shield}<0.2$; black lines) and high shielding efficiency ($\eta_\mathrm{shield}>0.2$; blue lines). }
\label{fig:grid_dependence_P_A_two_bins_eta_shield_paper_test02_mode_Q}
\end{figure*}

\begin{figure}
\center
\includegraphics[scale = 0.48, trim = 10mm 0mm 0mm 0mm]{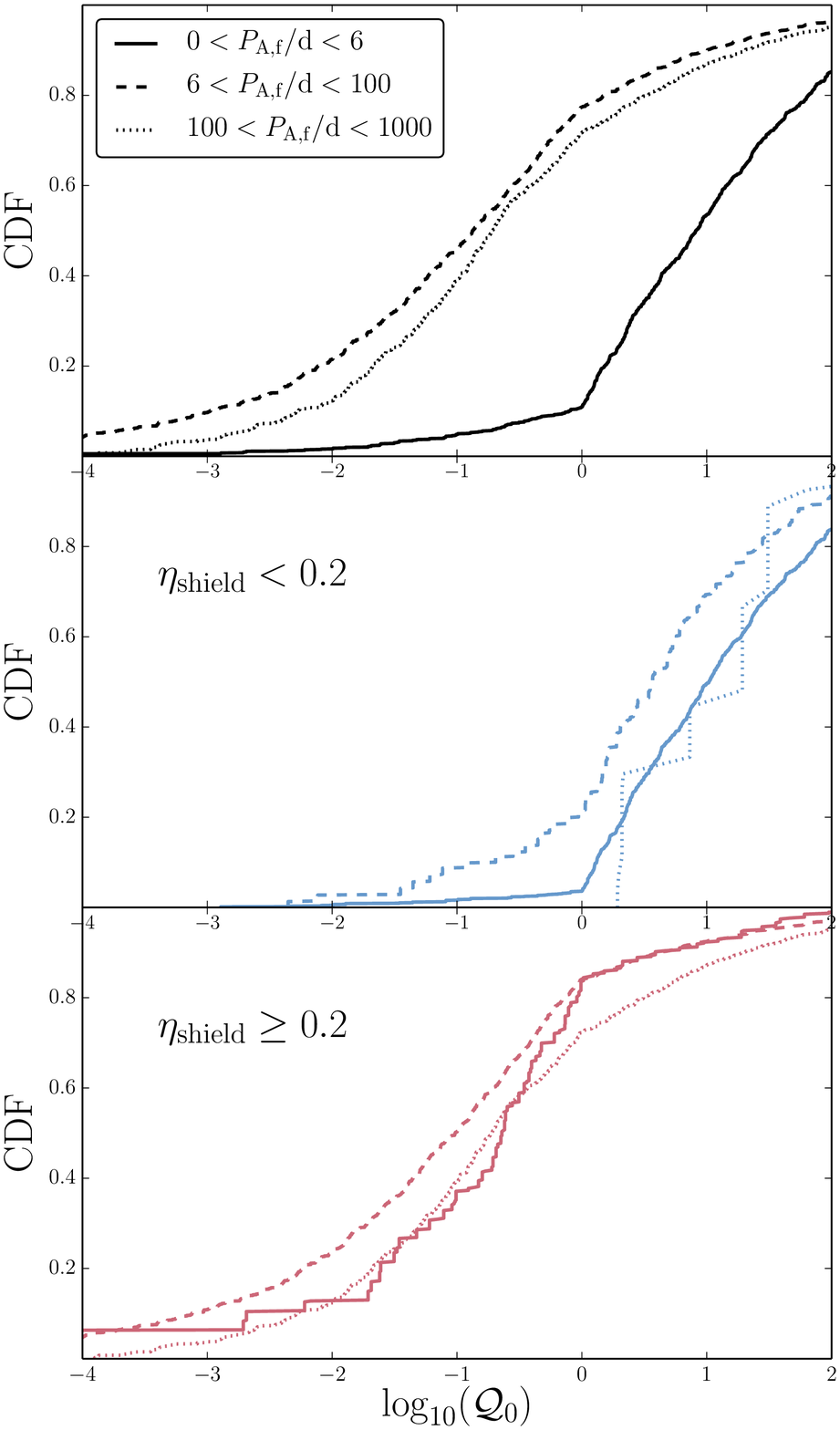}
\caption{\small Cumulative distributions for $\mathcal{Q}_0$, binned with respect to the final inner orbital period $P_\mathrm{A,f}$ (indicated in the legend in the top panel). In the middle (bottom) panels, a distinction is made with respect to the shielding efficiency $\eta_\mathrm{shield}$ (cf. equation~\ref{eq:eta_shield}): including the systems with $\eta_\mathrm{shield}<0.2$ (middle panel), and with $\eta_\mathrm{shield}>0.2$ (bottom panel). The meaning of the linestyles is the same between the different rows. }
\label{fig:grid_dependence_P_A_CDF_paper_test02_mode_Q}
\end{figure}

In Fig. \ref{fig:grid_dependence_P_A_two_bins_eta_shield_paper_test02_mode_Q} we show, for each of the final period bins of Fig. \ref{fig:grid_dependence_P_A_paper_test02_mode_Q}, the initial distributions of $\alpha_\mathrm{B}$, $\mathcal{Q}_0$, $i_\mathrm{B}$ and $m_3$, where additional binning was made with respect to $\eta_\mathrm{shield}$. Black (blue) lines correspond to $\eta_\mathrm{shield}<0.2$ ($\eta_\mathrm{shield}>0.2$). Cumulative distributions for the different final period bins and the two ranges of shielding efficiencies are shown in Fig. \ref{fig:grid_dependence_P_A_CDF_paper_test02_mode_Q}. As expected, the close planetary orbits from the top row in Fig. \ref{fig:grid_dependence_P_A_paper_test02_mode_Q} ($1 \, \mathrm{d} <P_\mathrm{A,f}<6\, \mathrm{d}$) typically correspond to a high shielding efficiency (cf. the blue lines in Fig. \ref{fig:grid_dependence_P_A_two_bins_eta_shield_paper_test02_mode_Q}). This indicates that the inner binary did not shrink, but was simply formed with a planet in such an orbit. On the other hand, the wider orbits typically correspond to a low shielding efficiency (cf. the black lines in Fig. \ref{fig:grid_dependence_P_A_two_bins_eta_shield_paper_test02_mode_Q}). In the latter case, the initial inner orbit was much wider, and shrank substantially. 

\subsubsection{Unstable planets}
\label{sect:pop_syn:results:unstable}
\begin{figure}
\center
\includegraphics[scale = 0.48, trim = 10mm 0mm 0mm 0mm]{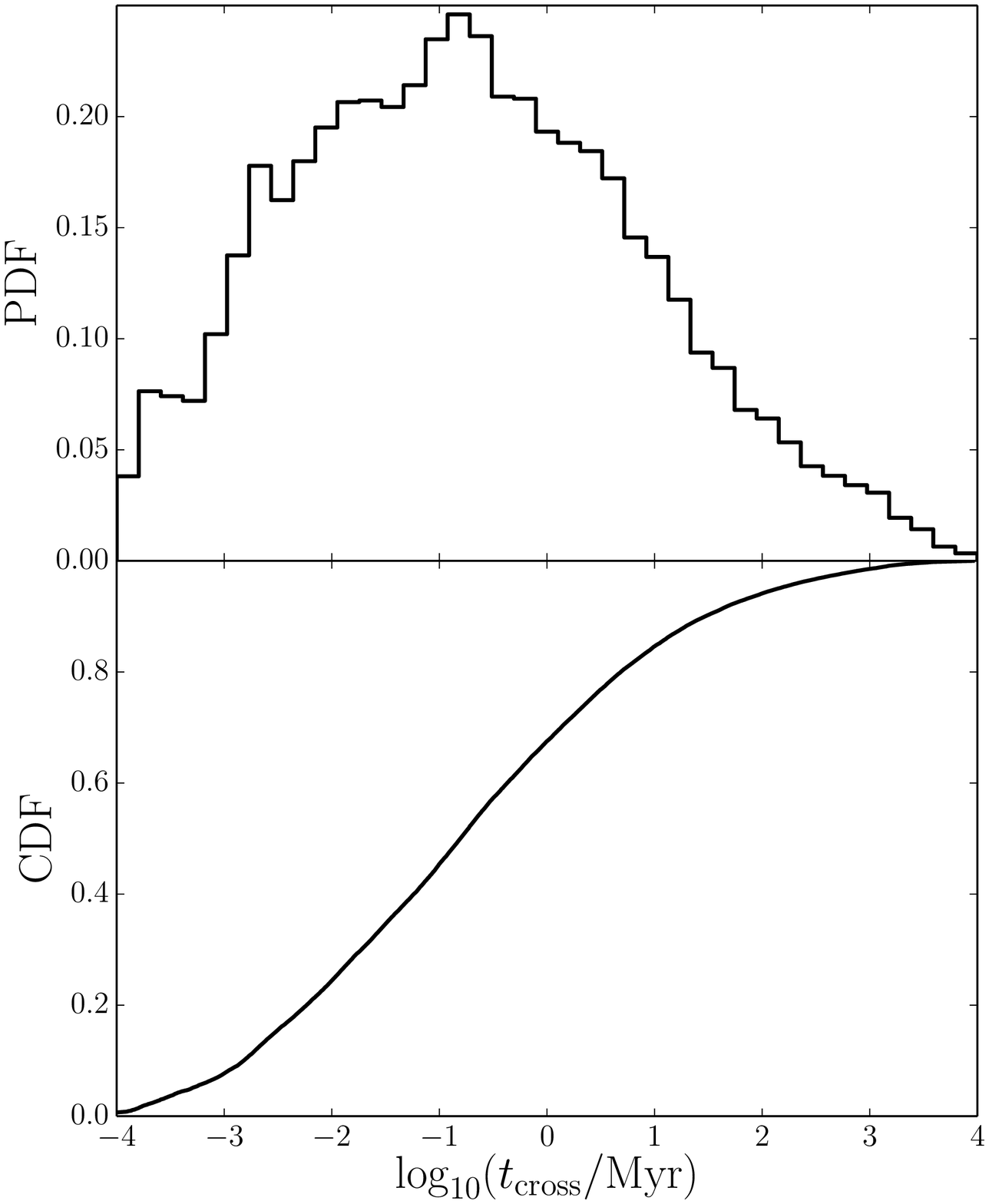}
\caption{\small The distribution of the time of orbit crossing $t_\mathrm{cross}$ of the planet with the inner binary, for the systems in which this occurred. }
\label{fig:grid_dependence_t_cross_paper_test02_mode_Q}
\end{figure}

As mentioned in \S\,\ref{sect:pop_syn:results:planets_short_period}, for a large fraction of systems, the planetary orbit becomes highly eccentric because of the torque of the outer orbit, and intersects with the inner binary. This likely results in an unstable planetary orbit, with a high probability for the planet being ejected from the system (for further discussion on and calculations of the outcomes, see \S\,\ref{sect:discussion:unstable}). Such orbit crossings typically occur early in the evolution. This is illustrated in Fig. \ref{fig:grid_dependence_t_cross_paper_test02_mode_Q}, where we plot the distribution of the time of orbit crossing $t_\mathrm{cross}$, when applicable. For $\approx 80\%$ of the cases, $t_\mathrm{cross}<10 \, \mathrm{Myr}$, which is very short compared to the MS time-scale of any of the stars (as described in \S\,\ref{sect:pop_syn:init_cond}, the integration time is sampled linearly between 1 and 10 Gyr). Because of this quick removal of the planet, it is unlikely that the planet could affect the inner binary evolution in these cases, and, therefore, we correspondingly assume $\eta_\mathrm{shield} = 0$. 

In Fig. \ref{fig:grid_dependence_output_fractions_paper_test02_mode_Q}, we show the fraction $f_\mathrm{cross}$ of systems with orbit crossing as a function of the initial parameters $\alpha_\mathrm{B}$, $\mathcal{Q}_0$, $i_\mathrm{B}$ and $m_3$. This fraction increases strongly from $f_\mathrm{cross}\approx 0.4$ for $\alpha_\mathrm{B}\approx0$ to $f_\mathrm{cross}\approx 0.9$ for $\alpha_\mathrm{B}\gtrsim 0.5$. This may seem counterintuitive: for larger $\alpha_\mathrm{B}$, the planet is further away from the inner binary, therefore orbit crossings are expected to be less likely. However, for larger $\alpha_\mathrm{B}$, the planet is also placed closer to the tertiary star, causing greater eccentricity excitation in the circumbinary orbit (e.g. the top-left panel in Fig. \ref{fig:maximum_eccentricity_test02}). Apparently, the latter effect is (much) stronger than the former, resulting in more orbit crossings for wider planetary orbits. As mentioned in \S\,\ref{sect:pop_syn:results:planets_short_period}, the high occurrence of orbit crossings produces a tendency for a lack of {\it stable} planets for $\alpha_\mathrm{B}\gtrsim0.3$.

There is no strong dependence of $f_\mathrm{cross}$ on $i_\mathrm{B}$, nor $m_3$. This can be understood by noting that the distribution of the mutual inclination between the circumbinary and the outer orbits, $i_\mathrm{BC}$, is the same as the distribution of the mutual inclination between the inner and outer orbits, $i_\mathrm{AC}$ (i.e. $\cos[i_\mathrm{AC}]$ was sampled linearly between -1 and 1; cf. \S\,\ref{sect:pop_syn:init_cond}). Therefore, the distribution of $i_\mathrm{BC}$ is independent of $i_\mathrm{B}$, and there is no preference with regard to $f_\mathrm{cross}$ for orbits with $i_\mathrm{B}$ close to $90^\circ$. Furthermore, because the planet mass $m_3$ is typically less than $10^{-3}$ of the stellar masses, $P_\mathrm{KL,BC} \propto (m_1+m_2+m_3+m_4)/m_4$ is nearly independent of $m_3$, explaining why $f_\mathrm{cross}$ is essentially independent of $m_3$. Note that, in contrast, the shielding efficiency $\eta_\mathrm{shield}$ is a strong function of $m_3$ (cf. Fig. \ref{fig:grid_dependence_eta_shield_paper_test02_mode_Q}). 

\subsubsection{Final planet orbital eccentricities}
\label{sect:pop_syn:results:planet_eccentricities}
\begin{figure}
\center
\includegraphics[scale = 0.42, trim = 10mm 0mm 0mm 0mm]{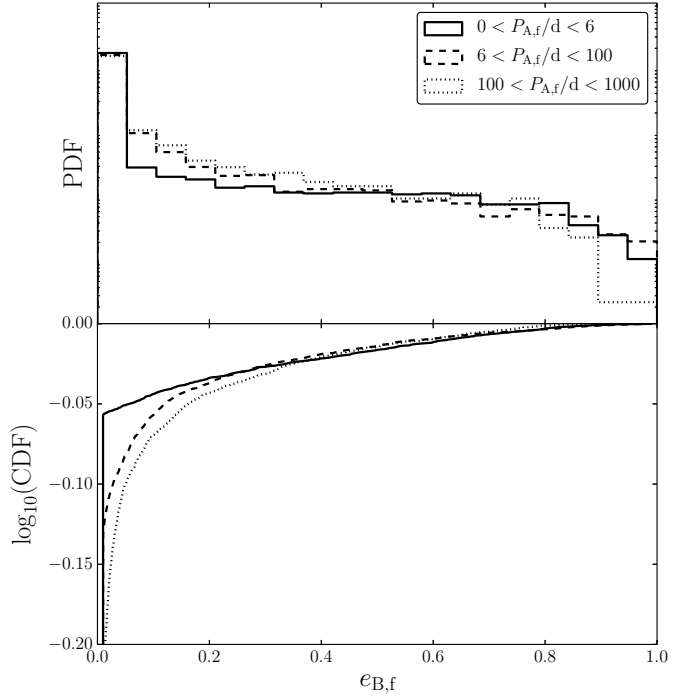}
\caption{\small The distributions of the final eccentricity  $e_\mathrm{B,f}$ of the planetary orbit for stable systems and binned with respect to the final inner orbital period $P_\mathrm{A,f}$. }
\label{fig:grid_dependence_P_A_eB_final_paper_test02_mode_Q}
\end{figure}

\begin{figure}
\center
\includegraphics[scale = 0.42, trim = 10mm 0mm 0mm 0mm]{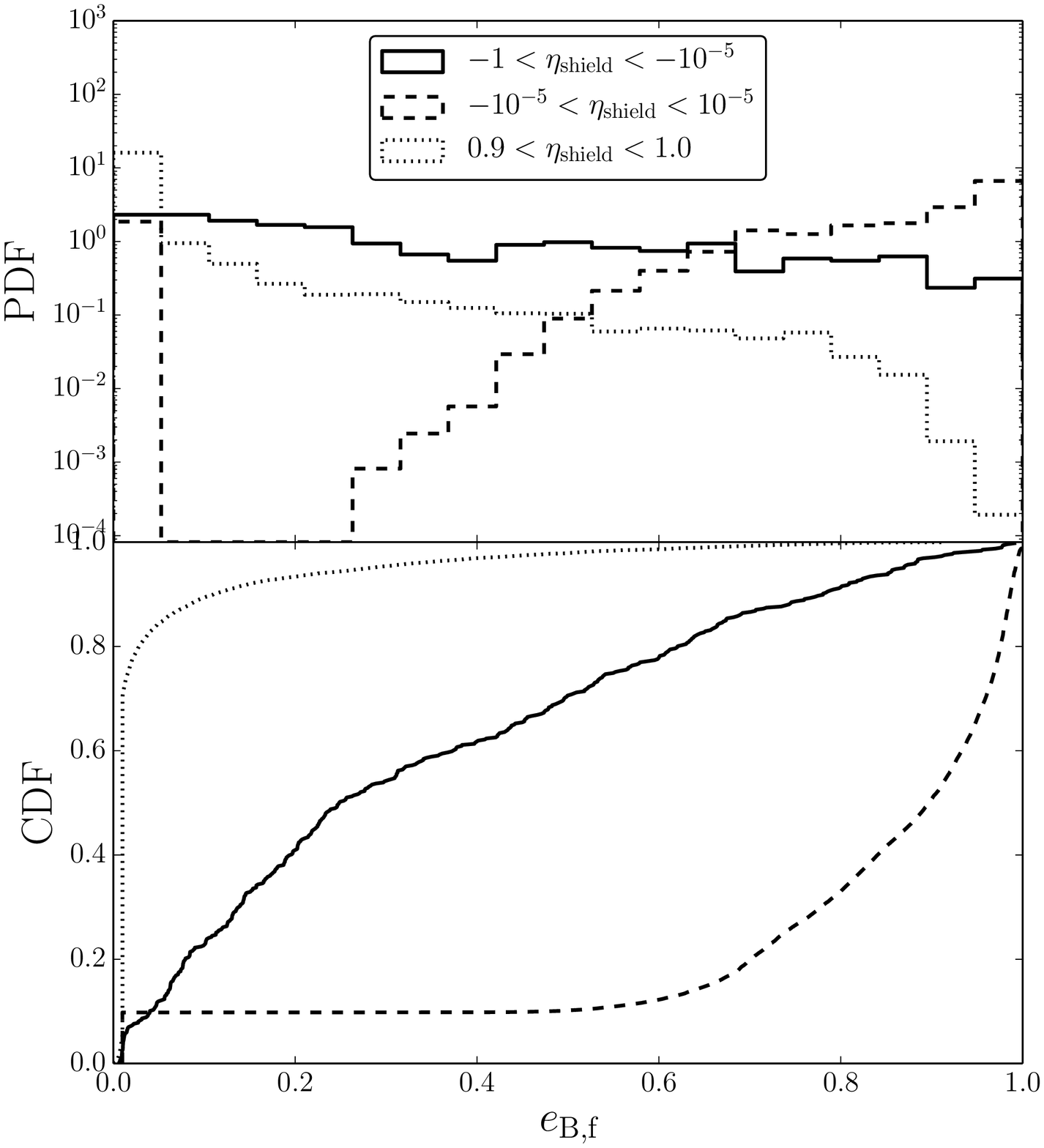}
\caption{\small The distributions of the final eccentricity  $e_\mathrm{B,f}$ of the planetary orbit, binned with respect to the shielding efficiency $\eta_\mathrm{shield}$. Here, cases are included when the planetary orbit intersects with the inner binary, in which case $\eta_\mathrm{shield}$ is assumed to be 0. }
\label{fig:grid_dependence_eta_shield_final_eB_paper_test02_mode_Q}
\end{figure}

In Figs \ref{fig:grid_dependence_P_A_eB_final_paper_test02_mode_Q} and \ref{fig:grid_dependence_eta_shield_final_eB_paper_test02_mode_Q}, we show the distributions of the planet orbital eccentricity at the end of the integration, $e_\mathrm{B,f}$, binned with respect to $P_\mathrm{A,f}$ and $\eta_\mathrm{shield}$, respectively. In the case of binning with respect to $\eta_\mathrm{shield}$, orbit crossing cases are also included, in which case we set $\eta_\mathrm{shield}=0$ (cf. \S\,\ref{sect:pop_syn:results:unstable}). For a crossing of the planetary orbit with the inner binary to occur, $e_\mathrm{B}$ needs to be high. This causes $e_\mathrm{B,f}$ to be typically high if $\eta_\mathrm{shield}\approx0$; in $\approx 50\%$ of the cases, $e_\mathrm{B,f}>0.9$. On the other hand, if the planetary orbit remains stable, $e_\mathrm{B}$ tends to be much smaller. For cases of very effective shielding, i.e. $0.9<\eta_\mathrm{shield}<1$, $\approx 90\%$ of the systems have $e_\mathrm{B,f}<0.2$. In Fig. \ref{fig:grid_dependence_P_A_eB_final_paper_test02_mode_Q}, the planetary orbit remains stable in all cases, and $e_\mathrm{B,f}$ tends to be low. There is little to no dependence of the distribution of $e_\mathrm{B,f}$ on the final orbital period. 

\subsubsection{Reversed shielding}
\label{sect:pop_syn:results:reversed_shielding}
As shown in the top row of Fig. \ref{fig:grid_dependence_eta_shield_paper_test02_mode_Q}, in a relatively small number of cases, $\eta_\mathrm{shield}<0$, i.e. the inner binary shrinks {\it more} compared to the situation without a planet. The values of $\alpha_\mathrm{B}$ for which this occurs are strongly peaked towards small values, and $\mathcal{Q}_0$ is peaked near $\mathcal{Q}_0 = 1$. For $\mathcal{Q}_0 \approx 1$, the KL time-scales for the AB and AC pairs are approximately equal, and, as similarly mentioned with respect to the quantity $\mathcal{R}_0$ by \quadpaper, this can give rise to chaotic dynamics and possible enhancement of the eccentricity in the inner binary, and hence a negative $\eta_\mathrm{shield}$. Although interesting from a theoretical dynamical point of view, negative $\eta_\mathrm{shield}$ only occur for a small number of systems, i.e. a fraction of 0.032 of the stable systems, and 0.0084 of all systems. 

\begin{figure*}
\center
\includegraphics[scale = 0.55, trim = 15mm 0mm 0mm 0mm]{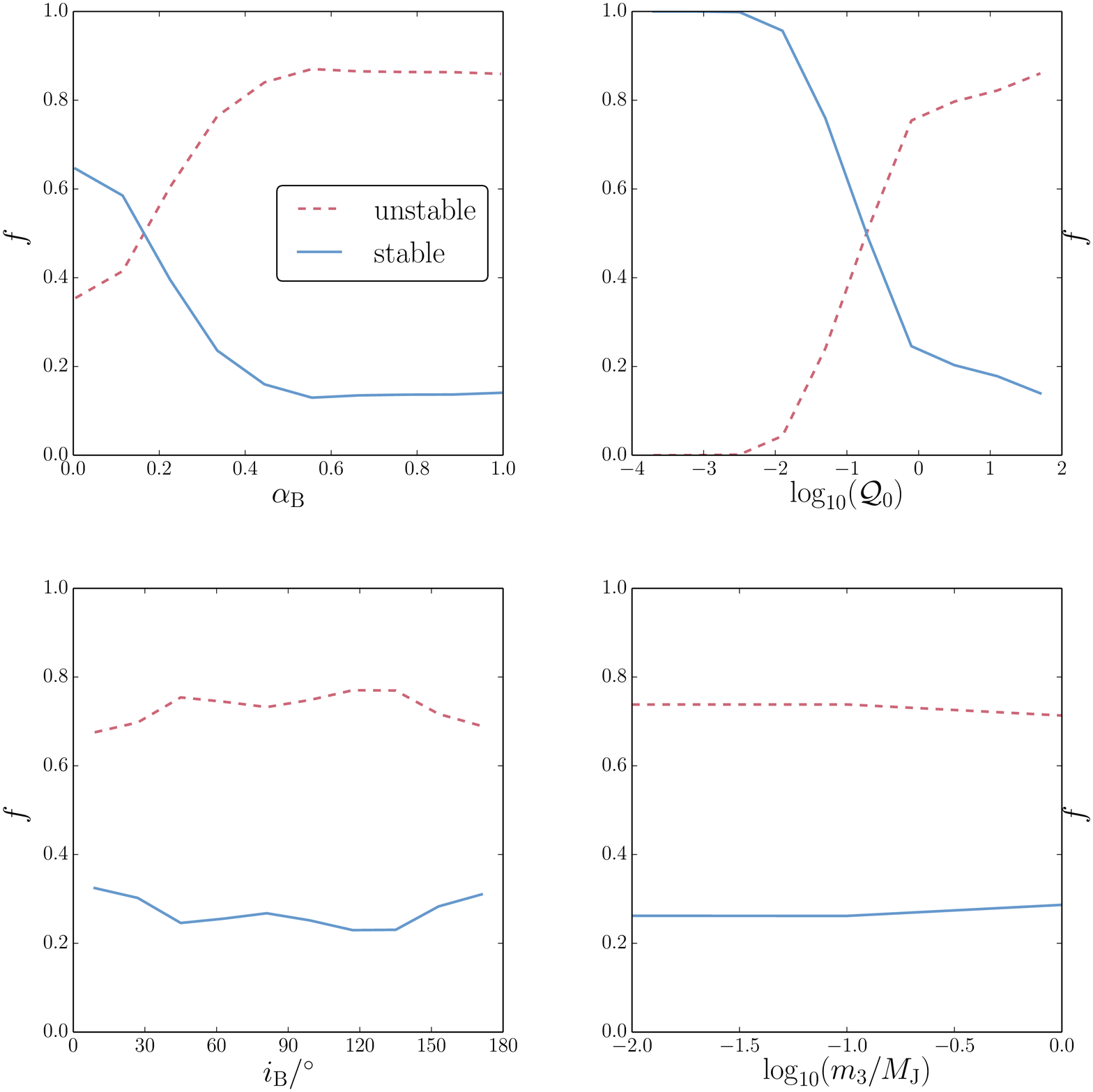}
\caption{\small The fractions of systems for which the planet either remains stable (blue solid lines), or for which the orbit of the planet becomes highly eccentric and intersects with the inner binary, likely resulting in an unstable orbit (red dashed lines), as a function of $\alpha_\mathrm{B}$, $\mathcal{Q}_0$, $i_\mathrm{B}$ and $m_3$. }
\label{fig:grid_dependence_output_fractions_paper_test02_mode_Q}
\end{figure*}

\section{Discussion}
\label{sect:discussion}

\subsection{An approximate analytic condition for planet shielding -- implications for other systems}
\label{sectl:discussion:condition}
In \S\,\ref{sect:shielding:kepler}, we have shown that planet shielding is effective for $\mathcal{Q}_0 \lesssim 1$ (cf. Fig. \ref{fig:maximum_eccentricity_test02}). We have also confirmed this for a much larger number of systems, and with the inclusion of tidal friction, in \S\,\ref{sect:pop_syn} (cf. Fig. \ref{fig:grid_dependence_eta_shield_paper_test02_mode_Q}). As demonstrated in e.g. Figs \ref{fig:maximum_eccentricity_test02}, \ref{fig:maximum_eccentricity_large_grid_test02ai3} and \ref{fig:grid_dependence_eta_shield_paper_subgroups_bins_test02_mode_Q}, there is also a dependence of the planet shielding ability on the inclination $i_\mathrm{B}$. Nevertheless, as a first approximation, the condition $\mathcal{Q}_0\lesssim 1$ can be used to evaluate the importance of planet-shielding in stellar triples. Using equation~(\ref{eq:Q0}), the condition can be written as
\begin{align}
\label{eq:Q0_cond_aC}
\nonumber a_\mathrm{B} &\lesssim a_\mathrm{C} \left ( \frac{m_3}{m_4} \frac{m_1+m_2+m_4}{m_1+m_2+m_3+m_4} \right )^{1/3} \left ( \frac{1-e_\mathrm{C,0}^2}{1-e_\mathrm{B,0}^2} \right )^{1/2} \\
\nonumber &\approx a_\mathrm{C} \left ( \frac{m_3}{m_4} \right )^{1/3} \left ( \frac{1-e_\mathrm{C,0}^2}{1-e_\mathrm{B,0}^2} \right )^{1/2} \\
&= 0.1 \, a_\mathrm{C} \left ( \frac{m_3}{m_4} 10^3 \right )^{1/3},
\end{align}
where in the second line, we assumed that $m_3$ is negligible compared to the other masses, and in the last line, we assumed $e_\mathrm{B,0}=e_\mathrm{C,0}=0$. For small planet masses, the scaling with $m_3$ is $a_\mathrm{B} \propto m_3^{1/3}$. Equation~\ref{eq:Q0_cond_aC} implies that for a planet of order Jupiter mass in a {\it stellar} triple (i.e. $m_k\sim 1 \, \mathrm{M}_\odot$ for $k\in \{1,2,4\}$), the circumbinary semimajor axis should be less than approximately a tenth of the outer semimajor axis. 

We remark that the fourth body orbiting the circumbinary planet system may also be less massive (by factors up to a thousand) than was assumed in \S s\,\ref{sect:shielding} and \ref{sect:pop_syn}. In particular, it could be a brown dwarf or a massive planet orbiting a lower-mass circumbinary planet. As $m_4$ decreases, equation~(\ref{eq:Q0_cond_aC}) implies that for fixed $a_\mathrm{C}$, $a_\mathrm{B}$ can be larger for shielding to be effective. This can be explained intuitively by noting that for fixed $m_3$ and decreasing $m_4$, the $P_\mathrm{KL,AB}$ time-scale remains constant, whereas the $P_\mathrm{KL,AC}$ time-scale increases, i.e. the outermost body becomes less dominant. This, combined with the lower efficiency of a tertiary with a lower mass to shrink the inner orbit, suggests that the lack of circumbinary planets around short-period binaries is even more severe for triples with low-mass tertiary companions. Similarly, shielding is expected to be more effective for lower inner binary masses $m_1$ and $m_2$, suggesting a more severe lack of planets around low-mass short-period binaries. 

\subsection{Shielding of the planetary orbit by the inner binary}
\label{sect:discussion:shielding_other}
Although not discussed in detail here, shielding can also occur in the orbit of the planet: the inner binary can induce rapid precession in the planetary orbit, shielding the latter against high-amplitude KL eccentricity oscillations induced by the outer binary. This aspect was discussed in detail by \quadpaper, where it was shown that the approximate condition for shielding of the planetary orbit is $\mathcal{R}_0 \lesssim 1$, or (cf. equation~\ref{eq:R0})
\begin{align}
\nonumber a_\mathrm{B} &\lesssim a_\mathrm{A}^{1/3} a_\mathrm{C}^{2/3} \left ( \frac{m_3}{m_4} \right )^{2/9} \left ( \frac{m_1+m_2+m_3}{m_1+m_2} \right )^{1/9} \left ( \frac{1-e_\mathrm{C,0}^2}{1-e_\mathrm{B,0}^2} \right )^{1/3} \\
\nonumber &\approx 0.22 \, a_\mathrm{A}^{1/3} a_\mathrm{C}^{2/3}  \left( \frac{m_3}{m_4} 10^3 \right )^{2/9}, \\
&\approx 0.92 \, \mathrm{AU} \, \left ( \frac{a_\mathrm{A}}{0.2 \, \mathrm{AU}} \right )^{1/3} \left ( \frac{a_\mathrm{C}}{20 \, \mathrm{AU}} \right )^{2/3} \left( \frac{m_3}{m_4} 10^3 \right )^{2/9}
\end{align}
where in the second and last lines, we neglected $m_3$ compared to the other masses, and where we assumed circular orbits. In general, $\mathcal{R}_0 \gg \mathcal{Q}_0$ (cf. equation~\ref{eq:R0}), indicating that typically, shielding of the inner binary by the planet ($\mathcal{Q}_0 \lesssim 1$) is more likely than shielding of the planet by the binary ($\mathcal{R}_0 \lesssim 1$). 

We note that similar dynamics apply to satellites around the Pluto-Charon binary system in the Solar system. As shown by \citet{michaely_ea_15}, if the orbit of a satellite is close enough to the Pluto-Charon binary, then precession induced by the Pluto-Charon binary on the satellite's orbit protects the latter from KL oscillations induced by secular perturbations of the Sun.

\subsection{The fate of planets with unstable orbits}
\label{sect:discussion:unstable}
As mentioned in \S s\,\ref{sect:shielding} and \ref{sect:pop_syn}, when the orbit of the planet is relatively close to the outer binary, the former can become highly eccentric because of KL eccentricity oscillations induced by the tertiary companion, and intersect with the inner binary. In fact, this is a likely scenario, as demonstrated by e.g. Fig. \ref{fig:grid_dependence_output_fractions_paper_test02_mode_Q}. Here, we investigate the possible outcomes of such orbit crossings and hence likely unstable orbits, by integrating the four-body system using the direct $N$-body code \textsc{Hermite} \citep{hut_ea_95} incorporated in \textsc{AMUSE} \citep{portegies_zwart_ea_13,pelupessy_ea_13}. Here, tidal effects are not included. The initial conditions are taken from the grid, of size 300, of the initial parameters $a_\mathrm{B}$, $i_\mathrm{B}$ and $m_3$ for one the triple systems sampled in \S\,\ref{sect:pop_syn}. Details of the parameters are given in the third column of Table \ref{table:init_cond1}. 

First, the systems are integrated using our secular code until the circumbinary orbit intersects with the inner binary. Subsequently, we sample 100 different sets of random mean anomalies for the three orbits, and integrate the system for each set for the duration of 40 initial circumbinary orbital periods. In this manner, the most important outcomes are revealed for each combination of planet parameters. In total, $300\times100=3 \times10^4$ direct $N$-body integrations were carried out.

We find the following outcomes, in order of decreasing likelihood $f$:
\begin{enumerate}
\item the planet becomes unbound from the system ($f\approx0.657$); \\
\item the planet orbits the outer binary (i.e. a circumtriple planet; $f\approx 0.289$) \\
\item the planet collides with a star ($f\approx0.029$); \\
\item the planet remains stable as a circumbinary planet, but with a different orbit ($f\approx0.017$); \\
\item the planet becomes bound to a single star ($f\approx0.009$). 
\end{enumerate}
The large probability of ejection of the planet is intuitively easy to understand from the low mass of the planet compared to that of the stars. Circumtriple planets are also common, but note that the integration time is limited and the orbits are typically eccentric, therefore not all systems may be stable indefinitely. 

\begin{figure}
\center
\includegraphics[scale = 0.48, trim = 10mm 0mm 0mm 0mm]{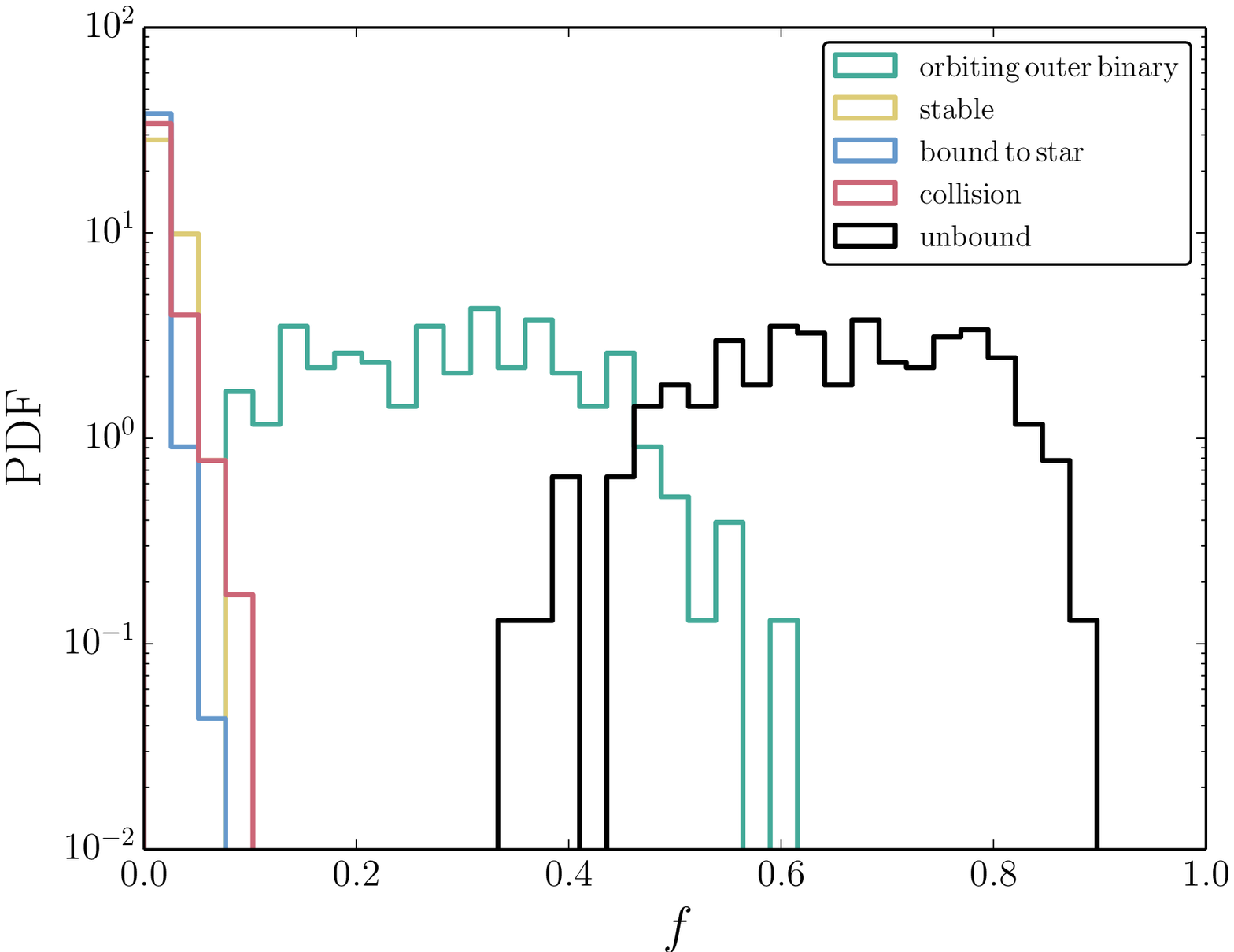}
\includegraphics[scale = 0.48, trim = 10mm 0mm 0mm 0mm]{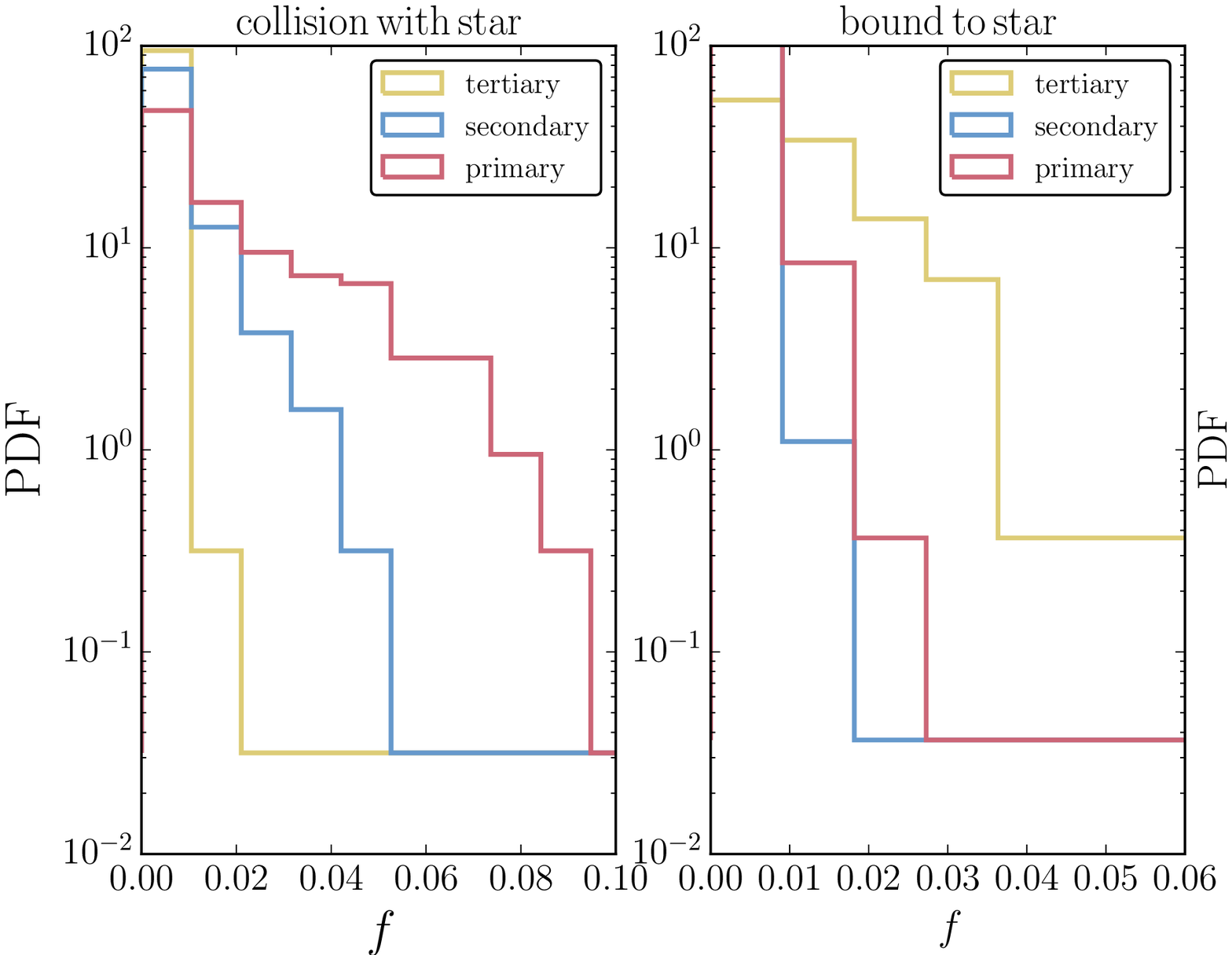}
\caption{\small The distributions of the fraction of direct $N$-body integrations with various outcomes, in the case that the orbit of the planet becomes highly eccentric and intersects with the inner binary. The triple system is fixed, whereas $a_\mathrm{B}$, $i_\mathrm{B}$ and $m_3$ are taken from a grid with size 300 as in \S\,\ref{sect:pop_syn:init_cond}. Top row: overview, with the major channels described in the legend. Bottom row: making a distinction between stars for cases when the planet collides with a star (left column), or when it becomes bound to a star (right column). }
\label{fig:nbody_check_dynamical_stability}
\end{figure}

In the top row of Fig. \ref{fig:nbody_check_dynamical_stability}, we show the distribution of the fractions of the outcomes. In the bottom row of the same figure, we show the distributions for collisions with stars (left column) and bindings to stars (right column), where a distinction is made between the different stars. Collisions are most common with the primary star. This is easily understood by noting that by definition, the primary star is the most massive, and, therefore, also has the largest radius (as described in \S\,\ref{sect:methods}, the radii are calculated using the \textsc{SeBa} stellar evolution code). With regards to the planet becoming bound to a single star, being bound to the tertiary is most likely. This may be the result of a lower orbital speed of the planet when it is close to the tertiary (roughly corresponding to the apocentre of the circumbinary orbit if it were still stable), as opposed to when it is close to the inner binary (roughly corresponding to the pericentre of the circumbinary orbit if it were still stable).

The outcomes found above -- keeping in mind the caveat that they are based on integrations of only a single triple system and that the planetary parameters were taken from a linear grid -- have interesting implications for stellar triples. Case (i), combined with the high fraction of orbit-crossing planets as found in \S\,\ref{sect:pop_syn:results:unstable}, suggests that circumbinary planets in triples are likely to become unbound from their parent binary early in their evolution, and become free-floating planets. Case (ii) suggests that circumtriple planets should be fairly common. Case (iii) provides a scenario for polluting stars in the inner binary, the primary star in particular. Lastly, case (v) suggests that in stellar triples, circum{\it binary} planets (i.e. P-type planets) can be transformed into circum{\it stellar} planets (i.e. S-type planets), albeit with a low probability.

\subsection{Implications for planets around blue straggler stars}
\label{sect:discussion:blue_stragglers}
As first suggested by \citet{perets_fabrycky_09}, and later studied more quantitatively by \citet{naoz_fabrycky_14}, KLCTF may also lead to mergers or mass transfer in short-period binaries, producing blue straggler stars (BSSs). Our results suggest (secular dynamical) constraints for planets around BSSs formed in this manner, where the planets are either in a circumbinary configuration (in case of binary BSSs) or in a circumstellar configuration (in case when the original stellar binary merged to produce a single BSS star). Analogously to the case of short-period binaries, we expect such planets around BSSs to be typically of low mass, and in wide and/or inclined orbits around the BSS or BSS binary. 

\subsection{Approximations in the integrations}
\label{sect:discussion:approximations}
In the numerical integrations of \S s\,\ref{sect:shielding} and \ref{sect:pop_syn}, the `cross' term that appears in the Hamiltonian at the octupole-order, $\overline{H}_\mathrm{oct,cross}$ (cf. section 2.4 of \quadpaper), was neglected. Note that the `non-cross terms', $\overline{H}_\mathrm{oct,AB}$, $\overline{H}_\mathrm{oct,BC}$ and $\overline{H}_\mathrm{oct,AC}$, were always included in \S s\,\ref{sect:shielding} and \ref{sect:pop_syn}. For the systems considered here, it is unlikely that the cross term $\overline{H}_\mathrm{oct,cross}$ has a large effect on the dynamical evolution because its numerical value is generally very small compared to the other terms that appear in the octupole and the next higher, hexadecupole, orders. For example, for the system chosen in \S\,\ref{sect:shielding:kepler} (cf. Table \ref{table:init_cond1}; setting $m_3=1\, M_\mathrm{J}$, $i_\mathrm{B}=0^\circ$ and $i_\mathrm{C} = 90^\circ$), the ratio $r$ of the absolute value of the orbit-averaged cross term to the absolute value of all other orbit-averaged terms at octupole and hexadecupole order, defined in equation (10) of \quadpaper, is $r \approx 6 \times 10^{-8}$ if $a_\mathrm{B} = 1 \, \mathrm{AU}$, and $r \approx 7 \times 10^{-7}$ if $a_\mathrm{B} = 5 \, \mathrm{AU}$. For the system of \S\,\ref{sect:shielding:tides} (setting $m_3=1\, M_\mathrm{J}$, $i_\mathrm{B}=0^\circ$ and $i_\mathrm{C} = 90^\circ$), $r \approx 6 \times 10^{-7}$ if $a_\mathrm{B} = 10 \, \mathrm{AU}$, and $r \approx 2 \times 10^{-7}$ if $a_\mathrm{B} = 50 \, \mathrm{AU}$.

Furthermore, in the integrations we assumed that there is only one planet orbiting the inner binary. Among the currently limited number of {\it Kepler} transiting circumbinary planets, there is already one system, Kepler 47 (cf. Table \ref{table:observed_systems}) with three confirmed circumbinary planets orbiting a single binary. Although the case of multiple planets is beyond the scope of this work and left for future work, we note that in this situation, the planet with the tightest orbit with respect to the inner binary likely has the largest potential for shielding the inner binary from the secular torque of the outer orbit. This picture is complicated by the possibility of planet-planet scattering and mean motion resonances if the planets are in close orbits to each other. 

\subsection{Other dissipative effects}
\label{sect:discussion:other_dis}
In this work, we considered the dissipative effect of tidal friction in the inner binary. Other potentially important dissipative processes that affect the orbital energies are gravitational wave (GW) emission, MB and gas drag in circumbinary discs. 

\subsubsection{Gravitational wave emission}
\label{sect:discussion:other_dis_GW}
Shrinkage due to GW emission was not included for numerical reasons, but we note that it is important only in very tight binaries that are beyond the scope of this paper. For circular binaries, the GW inspiral time is $t_\mathrm{GW} = 5 c^5 a_\mathrm{A}^5/[256 G^3 m_1 m_2 (m_1+m_2)]$, where $c$ is the speed of light \citep{peters_64}, which is equal to 10 Gyr if $\log_{10}(P_\mathrm{A}/\mathrm{d}) \approx -0.4$. In our population synthesis, the final orbital periods are typically longer than this value (cf. Fig. \ref{fig:period_distributions_test02_mode_B_and_T}), indicating that shrinkage in the inner orbit due to GW emission is not important (note that the inner orbit is typically nearly circular when the final orbital period is reached). 

\subsubsection{Magnetic braking}
\label{sect:discussion:other_dis_MB}
Many uncertainties still exist regarding the efficiency of MB. We evaluated the effect of MB by carrying out the integrations of isolated binaries and triples as in \S\,\ref{sect:pop_syn:init_cond}, now also including MB for the primary and secondary stars in the inner binary, with standard assumptions for the MB law and its efficiency. Further details are given in Appendix \ref{app:MB}. 

We find that when including the effects of MB, there is a nearly absolute absence of binaries with periods $\lesssim 10\, \mathrm{d}$ (cf. Fig. \ref{fig:period_distributions_test02MB_mode_B_and_T}), which is grossly inconsistent with observations. The latter, in contrast, show a {\it peak} in the distribution around 3-6 d \citep{tokovinin_14a}. This suggests that MB is likely not as efficient as was assumed here. Because MB is not the focus of this work, we chose to exclude it in the integrations of \S s\,\ref{sect:shielding} and \ref{sect:pop_syn}. Nevertheless, the apparent discrepancy between the canonical efficiency of MB and the period distribution of solar-type MS binaries merits future investigation.

\subsubsection{Drag in circumbinary gas discs}
If there is a disc around the inner binary, gas drag can cause the circumbinary planet to migrate inwards. Consequently, $\mathcal{Q}_0$ would decrease (cf. equation~\ref{eq:Q0}), and hence the planet's shielding ability would be increased. However, the lifetime of circumbinary discs is at most a few Myr \citep{alexander_12}, which is much shorter than the typical duration of our integrations (in our population synthesis, the integration time varied between 1 and 10 Gyr to reflect the typical age of observed {\it Kepler}-like systems). Therefore, disc migration likely does not affect our conclusions regarding planet shielding in {\it Kepler}-like systems.

\section{Conclusions}
\label{sect:conclusions}
In recent observations of circumbinary planets, there is an unexpected lack of coplanar circumbinary planets around short-period solar-type MS binaries. This goes {\it against} observational biases of detecting such planets. We have shown that the lack can be explained by the secular gravitational influence of a circumbinary planet in hierarchical triple systems. Our arguments and conclusions are as follows.

\medskip \noindent 1. Observations show that binaries with periods of $\sim3-6 \, \mathrm{d}$ are nearly always (96\%) orbited by a tertiary companion star \citep{tokovinin_ea_06}. The short orbital period likely resulted from Kozai-Lidov (KL) eccentricity oscillations in the inner orbit induced by the tertiary, combined with tidal friction \citep{mazeh_shaham_79,eggleton_ea_01,fabrycky_tremaine_07,naoz_fabrycky_14}. This suggests that many of the short-period {\it Kepler} eclipsing binaries are triple star systems. The progenitor inner binary was likely wider, with an orbital period of up to $\sim 10^4\, \mathrm{d}$. 

\medskip \noindent 2. We have demonstrated that if there is a circumbinary planet around the progenitor inner binary, then the KL eccentricity oscillations in the inner orbit induced by the tertiary can be quenched through the secular gravitational influence of the circumbinary planet. Thereby, the inner binary is `shielded' from the torque of the tertiary star, and does not shrink to a tight orbit. However, this only occurs if the circumbinary planet is sufficiently massive, and if its orbit is sufficiently close to and coplanar with the inner binary. In many other cases, the circumbinary orbit is stable but cannot efficiently shield the inner binary, or the circumbinary orbit is destabilized and is most likely ejected. 

\medskip \noindent 3. In particular, if a low-mass circumbinary planet is initially inclined with respect to and far away from the inner binary, then planet shielding is typically ineffective, and the inner binary can shrink to a tight orbit. On the other hand, for a more massive planet in an initially approximately coplanar and tight circumbinary orbit, shielding typically prevents the inner orbit from shrinking (cf. Figs \ref{fig:grid_dependence_P_A_two_bins_eta_shield_paper_test02_mode_Q} and \ref{fig:grid_dependence_eta_shield_paper_test02_mode_Q}). 

\medskip \noindent 4. Consequently, for systems with circumbinary planets surviving through the evolution, short-period inner binaries typically do not have massive ($m_3\sim M_\mathrm{J}$) circumbinary planets on tight ($\mathcal{Q}_0\lesssim 1$, where $\mathcal{Q}_0$ is defined in equation~\ref{eq:Q0}) and coplanar orbits. Namely, if the latter was present initially, the inner binary would unlikely have shrunk to a short period. In contrast, if the circumbinary planet is less massive ($m_3\sim 10^{-2} \, M_\mathrm{J}$) and initially inclined with respect to and far away from the inner binary ($\mathcal{Q}_0\gtrsim 1$), then shrinkage is possible. This trend is consistent with the current {\it Kepler} observations. The transition with respect to the planet mass and orbital semimajor axis occurs at $\mathcal{Q}_0 \sim 1$, which, for a Jupiter-mass planet in a solar-mass triple, corresponds to a circumbinary semimajor axis roughly equal to a tenth of the outer orbit semimajor axis (cf. equation~\ref{eq:Q0_cond_aC}). 

Our results suggest that similar constraints also apply to planets around blue straggler stars.

\section*{Acknowledgements}
We thank the referee, Smadar Naoz, for providing very helpful comments that led to improvement of the paper. This work was supported by the Netherlands Research Council NWO (grants \#639.073.803 [VICI],  \#614.061.608 [AMUSE] and \#612.071.305 [LGM]) and the Netherlands Research School for Astronomy (NOVA). HBP acknowledges support from the ISF I-CORE programme 1829, The European FP-7 CIG programme `GRAND' (333644), the BSF grant number 2012384 and the Asher foundation.

\bibliographystyle{mnras}
\bibliography{literature}

\appendix
\newpage

\section{{\it Kepler} transiting circumbinary planets}
\label{app:kepler}
In Table \ref{table:observed_systems}, we give an overview of the currently known {\it Kepler} transiting circumbinary planets. 

\begin{table*}
\begin{threeparttable}
\begin{tabular}{lccccccccccccc}
\toprule
System & $m_1$ & $m_2$ & $m_3$ & $R_1$ & $R_2$ & $R_3$ & $a_\mathrm{A}$ & $P_\mathrm{A}$ & $e_\mathrm{A}$ & $a_\mathrm{B}$ & $P_\mathrm{A}$ & $e_\mathrm{B}$ & Reference \\
 & $\mathrm{M}_\odot$ & $\mathrm{M}_\odot$ & $M_\mathrm{J}$ & $\mathrm{R}_\odot$ & $\mathrm{R}_\odot$ & $R_\mathrm{J}$ & AU & d & - & AU & d & - & \\
\midrule
Kepler 16b     & 0.690   & 0.203     & 0.333            & 0.649     & 0.226     & 0.754         & 0.224     & 41.1  & 0.159     & 0.705     & 229   & 0.0069  & a   \\
Kepler 34b      & 1.05    & 1.02      & 0.220            & 1.16      & 1.09      & 0.764         & 0.229     & 27.8  & 0.521     & 1.09      & 289   & 0.182  & b   \\
Kepler 35b      & 0.888   & 0.809     & 0.127            & 1.03      & 0.786     & 0.728         & 0.176     & 20.7  & 0.142     & 0.603     & 131   & 0.042 & c \\
Kepler 38b     & 0.949   & 0.249     & $<$0.384         & 1.76      & 0.272     & 0.388         & 0.145     & 18.8  & 0.103     & 0.464     & 106   & $<$0.032  & d \\
Kepler 47b      & 1.04    & 0.362     &                  & 0.964     & 0.351     & 0.266         & 0.0836    & 7.45  & 0.0234    & 0.296     & 49.5  & $<$0.035  & e \\ 
Kepler 47c     & 1.04    & 0.362     &                  & 0.964     & 0.351     & 0.411         & 0.0836    & 7.45  & 0.0234    & 0.989     & 303   & $<$0.411 & e\\ 
Kepler 47d   & 1.04    & 0.362     &                  & 0.964     & 0.351     &               & 0.0836    & 7.45  & 0.0234    &  0.72     &  187.3     &           & f \\ 
Kepler 64b      & 1.47    & 0.37      &                  & 1.7       & 0.34      &  0.55           & 0.177     & 20.0  & 0.204     & 0.642     & 139   & 0.1       & g \\
Kepler 413b     & 0.820   & 0.542     & 0.211            & 0.776     & 0.484     & 0.387         & 0.101     & 10.1  & 0.0365    & 0.355     & 66.3  & 0.118     & h \\
KIC 963289      & 0.934   & 0.194     & $3\times10^{-4}$ & 0.833     & 0.214     & 0.549         & 0.185     & 27.3  & 0.0510    & 0.788     & 241   & 0.0379    & i \\
\midrule
ROXs 42B b & 0.89 & 0.36 & 6-14 & & & & & & & $\sim$140 & & & j,k \\
FW Tau b & 0.28 & 0.28 & 6-14 & & & & & & & $\sim$330 & & & j,k  \\
Ross 458 C & 0.6 & 0.09 & 6-12 & & & & & & & 150-800 & & & k \\
SR12 C & 1.05 & 0.5 & 12-15 & & & & & & & $\sim 1100$ & & & k,l \\
DP Leonis & 0.6 & 0.09 & 6.39 &&&& 0.027 & 0.062 & 0 & 8.2 & 28 & 0.39 & m \\

\bottomrule
\end{tabular}
\begin{tablenotes}
            \item[a] \citet{doyle_ea_11} 
            \item[b] \citet{welsh_ea_12}
            \item[c] \citet{welsh_ea_12}
            \item[d] \citet{orosz_ea_12a} 
            \item[e] \citet{orosz_ea_12b}
            \item[f] Orosz (in prep)
            \item[g] \citet{kostov_ea_13}
            \item[h] \citet{kostov_ea_14}
            \item[i] \citet{welsh_ea_14}
            \item[j] \citet{kraus_ea_14}
            \item[k] \citet{bowler_ea_14}
            \item[l] \citet{kuzuhara_ea_11}            
            \item[m] \citet{schwope_ea_02,qian_ea_10,beuermann_ea_11}
\end{tablenotes}
\caption{ Observed circumbinary planets. The binary primary and secondary masses (radii), $m_1$ ($R_1$) and $m_2$ ($R_2$), respectively, are expressed in $\mathrm{M}_\odot$ ($\mathrm{R}_\odot$); the circumbinary planet mass (radius), $m_3$ ($R_3$), is expressed in $M_\mathrm{J}$ ($R_\mathrm{J}$). The binary and circumbinary orbits are denoted with A and B, respectively. The semimajor axes, $a_\mathrm{A}$ and $a_\mathrm{B}$, are expressed in AU, and the orbital periods, $P_\mathrm{A}$ and $P_\mathrm{B}$, are expressed in days. Note that so far, three circumbinary planets have been detected around the binary in the Kepler 47 system. Kepler 64b is also known as KIC 4862625; the binary+planet system is orbited by another stellar binary. Top rows: {\it Kepler} transiting circumbinary planets. Bottom rows: circumbinary planets detected through other methods. }
\label{table:observed_systems}
\end{threeparttable}
\end{table*}

\begin{figure}
\center
\includegraphics[scale = 0.48, trim = 15mm 0mm 0mm 0mm]{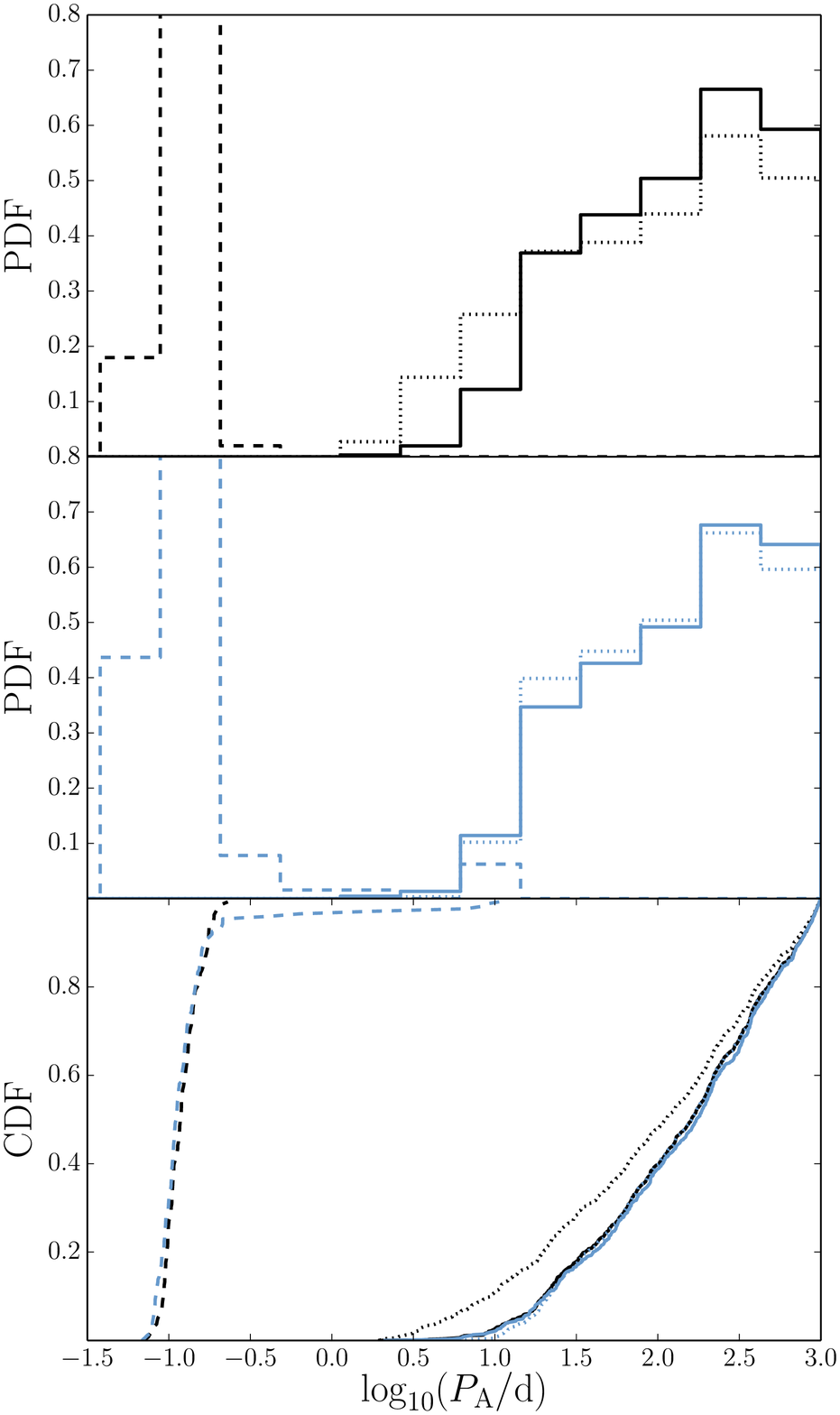}
\caption{\small The initial (black) and final (blue) inner period distributions for the sampled triples as described in \S\,\ref{sect:pop_syn:init_cond} (cf. Fig. \ref{fig:period_distributions_test02MB_mode_B_and_T}), here also including the effects of MB as discussed in \S\,\ref{sect:discussion:other_dis_MB} and Appendix \ref{app:MB}. Distributions corresponding to isolated binary (triple) evolution are shown with black (blue) lines. The initial distributions are shown with dotted lines. Cases when the inner binary merges are shown with dashed lines. Otherwise, the final distributions are shown with solid lines. }
\label{fig:period_distributions_test02MB_mode_B_and_T}
\end{figure}

\begin{figure}
\center
\includegraphics[scale = 0.42, trim = 10mm 0mm 0mm 0mm]{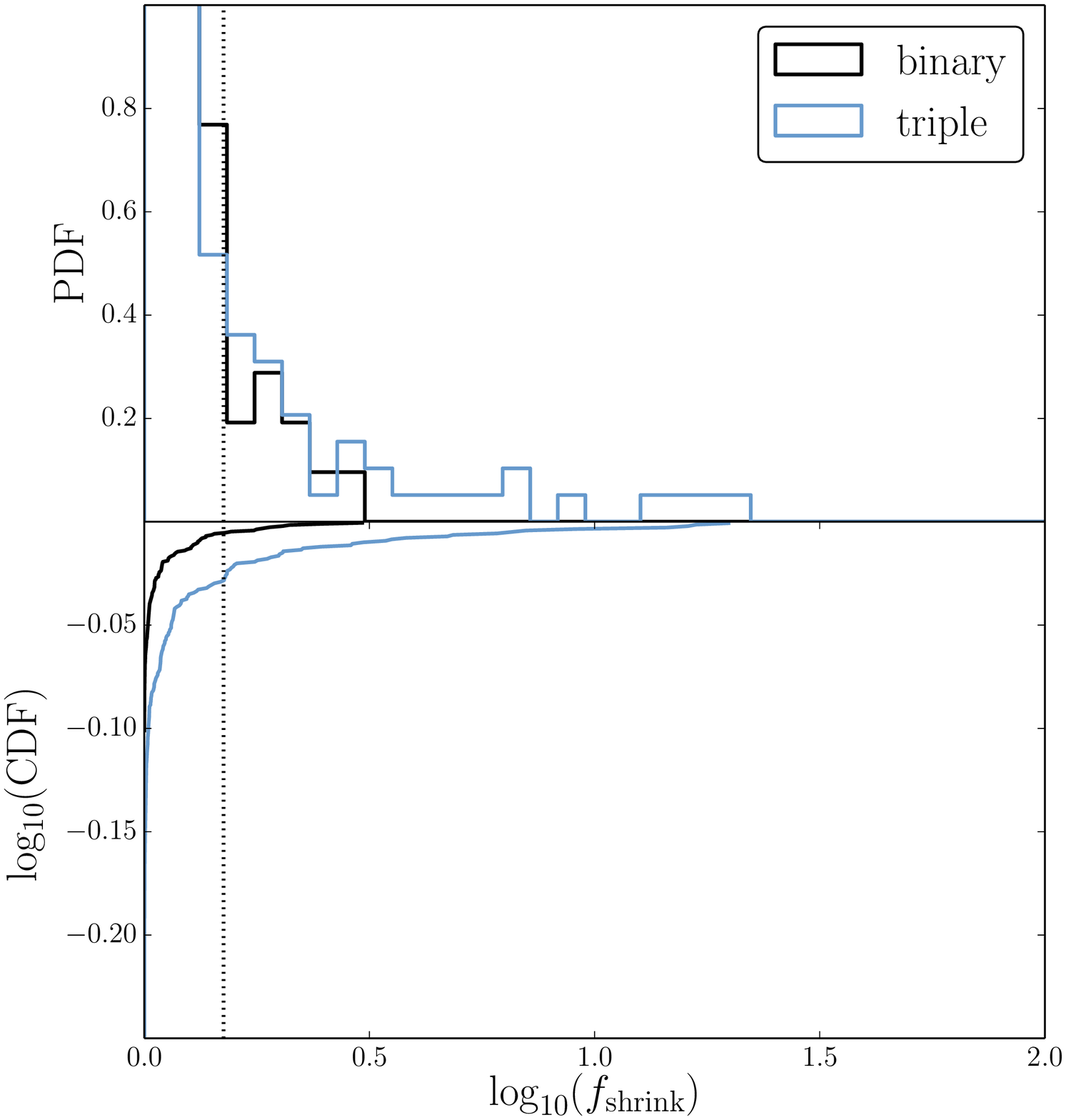}
\caption{\small The factor $f_\mathrm{shrink} \equiv a_\mathrm{A,i}/a_\mathrm{A,f}$ with which the inner binary period shrinks for the sampled triples as described in \S\,\ref{sect:pop_syn:init_cond}, here also including the effects of MB as discussed in \S\,\ref{sect:discussion:other_dis_MB} and Appendix \ref{app:MB}. Only systems are included for which the inner binary does not merge. Black (blue) lines: isolated binary (triple) evolution. The black vertical line indicates the cutoff value chosen for our population synthesis study (cf. \S\,\ref{sect:pop_syn:init_cond}). }
\label{fig:f_shrink_distributions_test02MB_mode_B_and_T}
\end{figure}

\section{Magnetic braking in triples}
\label{app:MB}
As discussed in \S\,\ref{sect:discussion:other_dis_MB}, we carried out the integrations of isolated binaries and triples as in \S\,\ref{sect:pop_syn:init_cond}, now also including the effects of MB for the primary and secondary stars in the inner binary by adopting equation (1) of \citet{barker_ogilvie_09}. The latter equation is based on the Skumanich relation \citep{skumanich_72} and the results of \citet{verbunt_zwaan_81,dobbs-dixon_ea_04}. Here, we assumed a MB parameter of $\alpha_\mathrm{MB}=1.5 \times 10^{-14} \, \mathrm{yr}$ \citep{barker_ogilvie_09}. 

In Fig. \ref{fig:period_distributions_test02MB_mode_B_and_T}, we show the distribution of the initial and final inner orbital periods as in Fig. \ref{fig:period_distributions_test02_mode_B_and_T}, but now with MB included. MB has a large effect on the orbital evolution. In contrast to the situation without MB, many systems merge during the evolution even for isolated binary evolution (cf. the dashed lines in Fig. \ref{fig:period_distributions_test02MB_mode_B_and_T}). Furthermore, when including the tertiary star, no peak is produced in the final period distribution for $P_\mathrm{A,f} \sim 1-6\, \mathrm{d}$ as in Fig. \ref{fig:period_distributions_test02_mode_B_and_T}. Instead, owing to the large number of merged systems, there is an {\it absence} of systems with these periods.

For the non-merging systems, we show the distributions of semimajor axis shrink factor $f_\mathrm{shrink}$ in Fig. \ref{fig:f_shrink_distributions_test02MB_mode_B_and_T}. Although the inner binary still shrinks more in the isolated triple case compared to the isolated binary case, the enhancement in $f_\mathrm{shrink}$ is not nearly as large compared to the situation without MB (cf. Fig. \ref{fig:f_shrink_distributions_test02_mode_B_and_T}). In other words, MB is so efficient at driving an inner binary merger in isolation, that KL cycles with tidal friction hardly affect the evolution.

As also discussed in \S\,\ref{sect:discussion:other_dis_MB}, this suggests that MB is likely not as efficient as was assumed here, and this apparent discrepancy between the canonical efficiency of MB and the period distribution of solar-type MS binaries merits future investigation.

\label{lastpage}
\end{document}